\documentclass[prodmode,acmtois]{acmsmall} % Aptara syntax

\usepackage[ruled]{algorithm2e}
\usepackage{graphicx}
\usepackage[round,authoryear]{natbib}
\usepackage{amsmath}
\usepackage{rotating}
\usepackage{fancyvrb}
\usepackage{amssymb}
\usepackage{tikz}
\usetikzlibrary{bayesnet}
\usepackage{pstricks-add}

\usepackage{amsthm}

\renewcommand{\algorithmcfname}{ALGORITHM}
\SetAlFnt{\small}
\SetAlCapFnt{\small}
\SetAlCapNameFnt{\small}
\SetAlCapHSkip{0pt}
\IncMargin{-\parindent}

\acmVolume{9}
\acmNumber{4}
\acmArticle{39}
\acmYear{2010}
\acmMonth{3}

\begin{document}

% Page heads
%\markboth{Ronan Cummins}{A P\'olya Urn Document Language Model for Improved Information Retrieval}

\title{A P\'olya Urn Document Language Model for Improved Information Retrieval}%\thanks{Grants or other notes


\author{Ronan Cummins
\affil{University of Cambridge}
Jiaul H. Paik
\affil{University of Maryland, College Park}
Yuanhua Lv
\affil{Microsoft Research}}

\begin{abstract}

The  multinomial language  model has  been one  of the  most effective
models  of  retrieval for  over  a  decade.  However, the  multinomial
distribution  does  not  model  one  important  linguistic  phenomenon
relating to term-dependency, that is  the tendency of a term to repeat
itself within a  document (i.e. word burstiness). In  this article, we
model document  generation as a  random process with  reinforcement (a
multivariate  P\'olya  process)   and  develop  a  Dirichlet  compound
multinomial language model that captures word burstiness directly.

We  show that the  new reinforced  language model  can be  computed as
efficiently as  current retrieval models,  and with experiments  on an
extensive  set of  TREC  collections, we  show  that it  significantly
outperforms  the  state-of-the-art  language  model for  a  number  of
standard effectiveness metrics. Experiments  also show that the tuning
parameter in the proposed model is more robust than in the multinomial
language model. Furthermore, we develop a constraint for the verbosity
hypothesis  and   show  that  the   proposed  model  adheres   to  the
constraint. Finally,  we show that the new  language model essentially
introduces a  measure closely related  to idf which  gives theoretical
justification  for combining  the term  and document  event  spaces in
tf-idf type schemes.

\end{abstract}

\category{H.3.3}{Information Storage and Retrieval}{Information Search and Retrieval}

\terms{Measurement, Theory, Performance}

\keywords{P\'olya Urn, Language Models, Smoothing, Retrieval Functions}

\acmformat{Ronan  Cummins, Jiaul  H.  Paik, and  Yuanhua  Lv, 2015.  A
  P\'olya  Urn  Document   Language  Model  for  Improved  Information
  Retrieval.}
% At a minimum you need to supply the author names, year and a title.
% IMPORTANT:
% Full first names whenever they are known, surname last, followed by a period.
% In the case of two authors, 'and' is placed between them.
% In the case of three or more authors, the serial comma is used, that is, all author names
% except the last one but including the penultimate author's name are followed by a comma,
% and then 'and' is placed before the final author's name.
% If only first and middle initials are known, then each initial
% is followed by a period and they are separated by a space.
% The remaining information (journal title, volume, article number, date, etc.) is 'auto-generated'.

\begin{bottomstuff}

Author's addresses: Ronan Cummins, The Computer Laboratory, University
of Cambridge;  Jiaul H. Paik,  College of Computer,  Mathematical, and
Natural  Sciences,  University  of  Maryland;  Yuanhua  Lv,  Microsoft
Research;
\end{bottomstuff}

\maketitle

\section{Introduction}

Language  modelling approaches  to information  retrieval  have become
increasingly      popular      since      the      original      works
\cite{ponte98,hiemstra98,hiemstra00,lavrenko01,zhai01}.  They afford a
particularly appealing  view of the  retrieval problem due in  part to
the  principled   nature  in  which   a  retrieval  function   can  be
mathematically derived. The  query likelihood method \cite{ponte98} is
one  of the  most widely-adopted  approaches to  retrieval,  and ranks
documents  based on the  likelihood of  their document  language model
generating  the  query string.  The  most widely-accepted  multinomial
language model treats the document model as a multinomial distribution
over  the terms,  where  the  parameters of  each  document model  are
estimated  using the  observations from  the actual  document smoothed
with the entire collection  using the Dirichlet prior smoothing method
\cite{zhai01}.

One  main  deficiency  with  using  a multinomial  distribution  as  a
language   model   is   that   all  term   occurrences   are   treated
independently.   The  term-independence   assumption   in  information
retrieval is  often adopted in theory  and practice as  it renders the
retrieval  problem tractable,  simplifies the  implementation  of many
models,  and  has  been  shown  to  be  suitably  effective.  Although
retrieval     approaches     that    incorporate     term-dependencies
\cite{metzler05,zhao09,lv09:02,cummins09,bendersky12}  have been shown
in  general  to  be  more  effective, they  are  computationally  more
complex. Therefore,  a language modelling  approach that has  the same
complexity  as   a  unigram  language  model   but  also  incorporates
dependencies,  would  be a  useful  contribution  as  it would  likely
exhibit  increased effectiveness  at no  extra computational  cost. In
fact, the use of the multinomial distribution in the standard language
modelling  approach ignores  two  types of  dependencies; namely,  the
dependency  between distinct  terms\footnote{This  is the  traditional
  term-independence  assumption.}  (word  types)  and  the  dependency
between recurrences of the same  term (word tokens). It is this second
type of dependency that we address in this article.

It is  well known that once  a term occurs  in a document, it  is more
likely to re-appear in the  same document. This phenomenon is known as
\emph{word  burstiness}  \cite{church95,madsen05}, and  is  a type  of
dependency  that is  not modelled  in the  multinomial  language model
\cite{zhai04}.  Essentially, word  burstiness  can be  defined as  the
tendency  of an  otherwise  rare term  to  occur multiple  times in  a
document, and can be seen  as a form of \emph{preferential attachment}
\cite{simon55,mitzenmacher03}. One theory  for this phenomenon is that
an  author  tends to  sample  terms  previously  written in  the  same
document  to form  \emph{association} \cite{simon55}.  The  process of
association of  similar concepts throughout a document  using the same
lexical form  may aid  coherence, readability, and  understanding. For
example, if  an author  starts to  use the term  {\bf pavement}  in an
article, he/she intuitively tends to continue its usage throughout the
document,  rather than  changing to  one  of its  synonyms (e.g.  {\bf
  sidewalk} or {\bf footpath}).

On  the other  hand,  queries  are requests  for  information and  are
generated  with  a  different  motive  in  mind.  When  requesting  or
searching  for  information  a  user  is more  likely  to  expand  the
vocabulary used  in the query (and  possibly make use  of synonyms) in
the  hope   of  matching  those  query-terms   contained  in  relevant
documents.  Furthermore,   queries  are  usually   much  shorter  than
documents and as  a result, we assume that queries  are less likely to
exhibit word-burstiness. That is not  to say that a certain term could
not  appear multiple times  in a  query, it  simply suggests  that the
reason for it  reappearing is different than in  a document. For these
reasons  we model  documents  and queries  using different  generative
assumptions.

This article  presents the {\bf SPUD} (Smoothed  P\'olya Urn Document)
language  model  that  incorporates  word  burstiness  only  into  the
document model. We use  the Dirichlet compound multinomial (DCM), also
known as the multivariate  P\'olya distribution, to model documents in
place  of the  standard  multinomial distribution,  while  we use  the
standard multinomial to model query  generation. We show that this new
retrieval model obtains significantly increased effectiveness compared
to the  current state-of-the-art  model on a  range of datasets  for a
number of  effectiveness metrics.  This article is organized as follows.
Section~2 introduces notation used in the remainder of the article and
also presents  a comprehensive review of  relevant research. Section~3
reviews the  standard language modelling  approach. Section~4 presents
the SPUD language model. Section~5 outlines efficient forms of the new
retrieval  functions, and  provides  deep insights  into the  proposed
functions.  The  experimental  design  and results  are  presented  in
Section~6.  Section~7  presents  a   discussion  of  the  results  and
Section~8 concludes with a summary.

\section{Related Research}

In this  section we  review related work  in language models  and word
burstiness,   before  outlining   the  main   contributions   of  this
work.  Table~\ref{tab:notation}   introduces  notation  used   in  the
remainder of this article.

\begin{table}[!ht]
%\scriptsize
\centering
\tbl{Feature Notation\label{tab:notation}}{

\renewcommand{\arraystretch}{1.5}
\setlength\tabcolsep{3pt}
\begin{tabular}{| l || l |}
\hline
Key	 	& Description \\
\hline
\hline
$c(t,d)$		& frequency of term $t$ in document $d$									\\	
$c(t,q)$		& frequency of term $t$ in query $q$										\\
$|d|$		& length of document $d$ (i.e. number of word tokens)					\\	
$\vec{|d|}$	& length of document vector (\# of distinct terms in document $d$	)		\\	
$cf_t$		& collection frequency (frequency of $t$ in the entire collection	)		\\	
$df_t$		& document frequency (number of documents in which $t$ occurs)			\\	
$|q|$		& length of query $q$ (i.e. number of word tokens)						\\
$|c|$		& number of tokens in the entire collection $c$							\\	
$n$			& number of documents in the collection 									\\
$|{v}|$		& vocabulary of the collection (\# of distinct terms in the collection) 	\\
\hline

\end{tabular}}
\end{table}

\subsection{Query Likelihood}

The  predominant  method  of  ranking  documents  using  the  language
modelling   approach   remains   the   query  likelihood   method   of
\citet{ponte98}. In the query  likelihood method, documents are ranked
based  on the  likelihood  of their  document model,  $\mathcal{M}_d$,
generating  the query  string. The  following equation  shows  how the
query  likelihood, $p(q|\mathcal{M}_d)$, is  calculated for  a unigram
multinomial language model:

\begin{equation} \label{eq:mql}
p(q|\mathcal{M}_d = \bm{\theta}_{dm}) = \prod_{t \in q} p(t | \bm{\theta}_{dm})^{c(t,q)}
\end{equation}
where  $q$   is  the  query  string  and   $\bm{\theta}_{dm}$  is  the
multinomial  document  language   model.  The  effectiveness  of  this
retrieval method  crucially depends on the estimation  of the document
model $\bm{\theta}_{dm}$.  It is typically estimated  using the actual
document $d$ and is smoothed  with the background language model which
is estimated from the entire collection $c$. When using a multinomial,
the query  likelihood method (Eq.~\ref{eq:mql}) can be  rewritten in a
rank equivalent form as follows:

\begin{equation} \label{eq:ql_log}
log\;p(q|\mathcal{M}_d = \bm{\theta}_{dm}) = \sum_{t \in q} (log \; p(t | \bm{\theta}_{dm}) \cdot {c(t,q)})
\end{equation}
which shows  that, as with  most other retrieval functions  (e.g. BM25
\cite{robertson96}),  the scoring  function comprises  a  summation of
query-term weights.  If $p(t  | \bm{\theta}_{dm})$ is  estimated using
only  the  maximum-likelihood  estimates  of  a term  occurring  in  a
document  (i.e.  $c(t,d)  /   |d|$),  over-fitting  would  occur.  For
instance,  this would  result in  any  document that  did not  contain
\emph{all}  query-terms not  being  retrieved, as  its document  model
deemed to have generated the query with a probability of zero
(see Eq.~\ref{eq:mql}). It should also be noted that when substituting
the   maximum   likelihood  probabilities   ($c(t,d)   /  |d|$)   into
Eq.~(\ref{eq:ql_log}), the weight of  each term becomes $log (c(t,d) /
|d|)$  which has  the effect  of reducing  the weight  contribution of
successive  occurrences of  the same  term to  a document  score. This
non-linear term-frequency  effect has been often reported  as a useful
heuristic                             in                            IR
\cite{fang04,fang05,cummins07,clinchant10,clinchantg11,
  lv12}. However, in the multinomial query likelihood retrieval method
this  non-linearity   is  only  the  consequence   of  a  mathematical
transformation,   and  the   actual   dependency  between   successive
occurrences of  the same term is  \emph{not} modelled\footnote{This is
  an important  point as  some previous work  tends to suggest  that a
  non-linear  term-frequency   factor  in  a   linear  combination  of
  term-weights   can   capture    some   aspect   of   dependency   or
  \emph{burstiness}. We contend that this is not the case for language
  models that use a multinomial as their basis.}.

\subsection{Advances in Language Models}

Since     the     initial     work    applying     language     models
\cite{ponte98,hiemstra98} to information  retrieval, there have been a
number of advances  in terms of both theory  and practice. Graph-based
models    \cite{gao04,metzler05,blanco12,bendersky12}   that   capture
aspects  of  term-dependency  have  been shown  to  improve  retrieval
performance   over  unigram   models.   Furthermore,  positional-based
language models \cite{zhao09,lv10}  have been proposed and incorporate
term  dependencies that  often  span several  terms.  In general,  the
incorporation of term-dependency information in larger web collections
has been shown to be beneficial to retrieval quality.

Although many  language modelling approaches  to information retrieval
use the query-likelihood approach to ranking, it is not the only means
of  inducing   a  ranking   using  language  models.   In  particular,
relevance-based language models \cite{lavrenko01} estimate a relevance
model from  which all relevant documents for  a particular information
need are assumed  to have been drawn. The approach  to ranking in that
work  is  similar  to  the classic  probabilistic  document  retrieval
approaches  \cite{jones00}, where  documents are  ranked based  on the
odds of being  drawn from the relevant class  compared to non-relevant
class.  The  relevance-based language  modelling  approach provides  a
principled mechanism  in which the  retrieval model can be  updated as
relevant and non-relevant documents become known. This approach led to
the  development of pseudo-relevance  query expansion  language models
\cite{umass4,diaz06,lv10}.

The language modelling  approach has now become a  starting point from
which more  complex models can  be built. Aside  from pseudo-relevance
query  expansion,  other  approaches  such as  \emph{latent  Dirichlet
  allocation}  (LDA)  have  been  incorporated into  ad-hoc  retrieval
\cite{wei06}. In essence, improving the retrieval effectiveness of the
standard  language  modelling approach  to  information retrieval  can
ultimately benefit any  of the myriad of approaches  which depend upon
it (e.g. pseudo-relevance feedback).

\subsection{Word Burstiness}

The  modelling of  word  burstiness in  documents  has been  addressed
before in  text-related tasks, but  it has not been  incorporated with
the query likelihood method in information retrieval. \citet{madsen05}
use the DCM distribution to  model word burstiness and demonstrate its
effectiveness on  document classification.  They estimate a  DCM model
for  each class  from  training  data. They  then  classify an  unseen
document to a specific class according to the mostly likely generative
DCM class model. They show that this DCM model outperforms the more standard
multinomial  model.   The  information  retrieval   task  is  somewhat
different as it deals with both  documents and queries. In our work we
have   different  generative  assumptions   for  both   documents  and
queries. A further difference is that in the classification task there
are a number  of documents from which we can  infer a particular class
model, while in the query-likelihood approach to information retrieval
we have  access to only one  instance of a document  from the document
model.

Due  to   the  complexity  of  estimating  parameters   for  the  DCM,
\citet{elkan06} developed  an approximate distribution  (the EDCM) and
demonstrated  its effectiveness for  clustering. We  make use  of this
approximation later in  this article. The DCM has also  been used in a
hypergeometric  language  model  \cite{tsagkias11} for  modelling  the
characteristics  of very  long  queries. In  other  work, a  two-stage
language modelling approach has been developed \cite{goldwater11} that
generates words according to  the power-law characteristics of natural
language.  They  decompose  the  language generation  process  into  a
\emph{generator},  which  creates  instances  of word  types,  and  an
\emph{adaptor} which  has the tendency  to repeat those  specific word
types.  Further arguments  which link  preferential attachment  to the
power-law  characteristics   of  natural  language   are  reviewed  by
\citet{mitzenmacher03}. \citet{cowans04} uses a hierarchical Dirichlet
process to  arrive at  a ranking function  which is reported  as being
superior  to   BM25.  Related  work   \cite{sunehag07}  provides  some
interesting  connections between  the  traditional {tf-idf}  weighting
scheme and  the two-stage generator-adaptor  models. Our work  is more
extensive and  actually develops a document language  model from which
retrieval functions are derived.

In recent  work, an extension of  earlier information-based approaches
\cite{amati02}  is   developed  that  incorporates   burstiness  in  a
log-logistic retrieval  function \cite{clinchantg09,clinchantg11}. The
authors  develop   a  means   for  identifying  if   a  term-frequency
distribution  is  \emph{bursty}.  They  conclude  that  the  frequency
distribution   must  be   a   type  of   power-law  (or   Pareto-type)
distribution.  Our work  is  much  more in  the  spirit of  generative
language  modelling where the  term-frequency aspect  occurs naturally
from  the model  (in our  case  a hierarchical  Bayesian approach)  to
introduce  dependencies  between subsequent  occurrences  of the  same
term.  Our model  also exhibits  power-law  characteristics consistent
with the work by \citet{clinchantg11}.

The work most  similar to ours uses the DCM  distribution to develop a
probabilistic  relevance-based  language  model \cite{xu08,xu10}.  For
each query they estimate a  relevant and non-relevant DCM model and it
is assumed that  all documents are generated from  either of those two
models.  However, our  work  does  not assume  a  relevance model  and
instead,  assumes that  each document  is generated  from  a different
document model. This means that  we model burstiness on a per document
basis, rather  than modelling  burstiness for a  set of  relevant (and
non-relevant) documents.  It is  more likely that  different documents
are  bursty to  different degrees  as they  were written  by different
authors, and this  is not modelled in the  relevance-based approach of
Xu  and  Akella.  Our  model  is  a  query-likelihood  approach  using
different generative assumptions for  both the document and query, and
leads  to retrieval  functions that  are  distinct from  those in  the
aforementioned relevance-based approach.

Although   \citet{xu08}   report   some   improvement   in   retrieval
effectiveness over  the multinomial query  likelihood retrieval method
on  some  test  collections,  their  experiments  were  restricted  to
relatively small collections (less  than a million documents) and used
only short keyword queries. It is unclear if their results extend to a
more general retrieval scenario. We  perform a more robust analysis by
using their best approach ({DCM-L-T})  as one of our main baselines on
a variety  of different  query lengths and  collection sizes.  We also
discuss the  difference between  our approach and  the relevance-based
DCM approach of Xu and Akella in our discussion section (Section~7.1).

\subsection{Contributions}

To our  knowledge no existing  work has developed a  document language
model  for  information  retrieval  using the  generative  assumptions
outlined  in this  work.  Therefore, the  main  contributions of  this
article are as follows:

\begin{list}{$\bullet$}{}  

\item We propose a new family of document language models that capture
  word burstiness in a probabilistic manner.

\item We  develop closed-form expressions for  the retrieval functions
  derived from  the new  language model, and  show that  our retrieval
  functions  are as  efficient as  traditional  bag-of-words retrieval
  functions.

\item  We show  that the  proposed language  model  implements several
  important  retrieval  heuristics  not  captured in  the  multinomial
  language model,  such as  modelling the \emph{scope  hypothesis} and
  the \emph{verbosity hypothesis} separately.

\item  We  show that  the  modelling of  word  burstiness  in the  new
  language  model  leads   to  significant  improvement  in  retrieval
  effectiveness for  ad hoc retrieval and for  downstream methods such
  as pseudo-relevance feedback.

\end{list}

We now  briefly review the  query likelihood retrieval method  and the
multinomial language model.

\section{Multinomial Language Model}

In this section we review  details of the multinomial query likelihood
model and some useful approaches to smoothing.

\subsection{Document and Background Models}

As outlined earlier,  it is the selection of  the generative model and
the  subsequent estimation  of  the document  language  model that  is
crucial  to   retrieval  effectiveness  using   the  query  likelihood
retrieval  method.  It  has  been  shown  \cite{zhai01,  zhai04}  that
effective  estimates of the  probability of  term occurrences  for the
multinomial document language model $\bm{\theta}_{dm}$ can be found as
follows:

\begin{equation}  \label{eq:ql_smoothing}
p(t| \bm{\hat{\theta}}_{dm}) = (1 - \pi) \cdot p(t|\bm{\hat{\theta}}_d) + \pi \cdot p(t|\bm{\hat{\theta}}_c)
\end{equation}
where  $\bm{\hat{\theta}}_{dm}$  is  the estimated  smoothed  document
language model and  $\pi$ is a smoothing parameter  which controls the
amount  of probability  mass  that should  be  redistributed from  the
background  multinomial  $p(t|\bm{\hat{\theta}}_c)$  to  the  document
multinomial $p(t|\bm{\hat{\theta}}_d)$.  This prevents over-fitting of
the  document  model  because  in  most  retrieval  formulations  both
$p(t|\bm{{\theta}}_d)$ and  $p(t|\bm{{\theta}}_c)$ are estimated using
maximum  likelihood   estimates  ${c(t,d)}/{|d|}$  and  ${cf_t}/{|c|}$
respectively.  The  background  multinomial  is  estimated  using  all
documents in  the entire  collection and therefore  all tokens  in the
corpus are  treated as independent observations.  The background model
can  be viewed  as  the most  likely  single model  to have  generated
\emph{all}  of the documents.  It has  been shown  that the  choice of
smoothing   greatly  affects  the   retrieval  effectiveness   of  the
multinomial language model \cite{zhai04}.

\subsection{Smoothing}

One of the simplest forms of smoothing uses linear-interpolation, also
called  Jelinek-Mercer smoothing,  where $\pi_{_{jm}}$  is  assigned a
value in  the range  $(0-1)$. In this  linear smoothing  approach, the
parameter  is usually  set by  experimentally tuning  $\pi_{_{jm}}$ on
training data. Typically there has  been no guidance on the setting of
this  parameter as  the effectiveness  of this  smoothing  approach is
quite  sensitive  to  specific   parameter  values.  However,  a  more
effective  smoothing method  for the  multinomial language  model uses
Bayesian smoothing in the form  of a Dirichlet prior on the background
multinomial. For this approach $\pi_{_{dir}}$ is defined as follows:

\begin{equation}  \label{eq:dir_smoothing}
\pi_{_{dir}} = \frac{\mu}{\mu + |d|}
\end{equation}
where  $\mu$ is  the concentration  parameter and  is the  sum  of the
individual $|v|$-Dirichlet parameters. This concentration parameter is
also assigned  a value  based on experimentation,  though it  has been
found that it achieves a relatively stable performance when $\mu=2000$
\cite{zhai04}. The Dirichlet prior  parameter $\mu$ can be interpreted
as the number of pseudo-counts  of the background multinomial prior to
the document data. Intuitively, this type of smoothing gives a greater
credence  to  probability  estimates  that  are  derived  from  longer
documents, compared  to those derived  from shorter documents,  as the
longer documents are likely to be more accurate representations of the
document   model.  The   prior  parameters   (pseudo-counts)   of  the
$|v|$-component  Dirichlet  distribution  are  $\alpha_t =  \mu  \cdot
p(t|\bm{\hat{\theta}}_c)$ for all $t \in  v$ and are updated using the
document     observations     to     $\alpha_t     =     \mu     \cdot
p(t|\bm{\hat{\theta}}_c) +  c(t,d)$ for all $t \in  v$. Therefore, the
concentration  parameter of  the Dirichlet  distribution  changes from
$\mu$ to $\mu+|d|$ once  the distribution has been updated. Throughout
this  article  we  will   continue  the  convention  of  specifying  a
$|v|$-component  Dirichlet  using  the  parameters  of  a  multinomial
distribution  (with  $|v|$-$1$ degrees  of  freedom)  multiplied by  a
concentration parameter (i.e. $\mu$).

It has been shown that the query likelihood model with Dirichlet prior
smoothing  and   the  model  with  Jelinek-Mercer   smoothing  can  be
implemented as  efficiently as traditional  retrieval functions, which
only  use weights  from terms  that are  common to  both  document and
query\footnote{See   the  original   source   \cite{zhai04}  for   the
  derivations of these efficient retrieval functions.}.

\section{A Smoothed P\'olya Urn Document Model}

In this section  we first introduce the generalised  P\'olya urn model
and outline  some of its  important characteristics. We then  show how
this can  be used to  model document generation before  specifying the
query likelihood approach  for the new model. Finally,  we outline how
the parameters of the SPUD model are estimated and smoothed.

\subsection{A P\'olya Urn Process}

Consider a  process that  starts with an  urn containing $m$  balls in
total, where each ball is one of $|{v}|$ distinct colours. Starting at
time $i=0$,  a ball is  sampled with replacement  from the urn,  and a
ball  of the  same colour  is replicated  and added  to the  urn. This
process  continues  until  $|d|$  balls  have been  sampled  from  the
urn. The total number of balls in the urn at the end of the process is
$m + |d|$.  This is a typical description  of the multivariate P\'olya
urn model which uses sampling  with reinforcement. We use this process
as  a conceptual model  for document  generation, where  the different
colours  represent distinct  terms, where  the initial  counts  of the
$|v|$  different coloured  balls  in the  urn  represent the  document
model,  and where the  $|d|$ observations  drawn represent  the actual
document.

This multivariate P\'olya urn model  has recently been described in an
alternative  manner as  consisting of  a multinomial  and  the Chinese
restaurant  process \cite{sunehag07,goldwater11}.  Again,  consider an
urn that contains  $m$ balls of $|v|$ different  colours, but now also
consider a bag $d$ that is  initially empty. For all times starting at
time $i=0$, a  ball is chosen from the  urn with probability $m/(m+i)$
and  from the  bag with  probability $i/(m+i)$,  and each  time  it is
replaced from  where it was drawn. For  each draw, a ball  of the same
colour that  was drawn  is generated  and placed in  the bag.  In this
alternative description,  the number of  balls $m$ in the  urn remains
static,  while the  number  of balls  in the  bag  $d$ is  $i$ at  any
particular  time.  The  non-reinforced   urn  can  be  modelled  as  a
multinomial  and the  bag can  be modelled  as the  Chinese restaurant
process.

This  two-stage  generative  process  has been  outlined  recently  by
\citet{goldwater11}  and  \citet{sunehag07},   and  while  the  entire
process is  identical to the multivariate P\'olya  urn model described
previous, it may  be more intuitive in terms of  a generative story of
document  creation. This  is because  the  document is  modelled as  a
separate entity  that starts  empty, and ends  after $|d|$  terms have
been drawn.  We re-introduce the alternative description  here only to
motivate  the  application  of   this  process  to  that  of  document
generation.  This is  very  much in  the  spirit of  that proposed  by
\citet{simon55} where an author  generates a document by drawing words
from some distribution and also by drawing words from those previously
used in the document in  order to create association. For the reminder
of the article, when we refer  to an \emph{urn}, we mean a P\'olya urn
by default, unless otherwise stated.

It is well-known that the  distribution of colours in the multivariate
P\'olya    process    follows    the   DCM    (multivariate    P\'olya
distribution). It is also known that  the P\'olya urn is an example of
a bounded martingale  process \cite{pemantle07}, where the proportions
of colours  in the urn  converges to a Dirichlet  distribution. During
the process,  the drawing, subsequent replication, and  addition of an
observation  (which must  be  identically distributed  to the  initial
distribution)    only     serves    to    reinforce     the    initial
distribution. Therefore,  all subsequent balls drawn  from the P\'olya
urn are identically distributed, but are not independent. Furthermore,
the process  is \emph{exchangeable}, meaning that the  ordering of the
outcomes   can  be  swapped   to  result   in  the   same  probability
distribution.  Therefore, the  document model  remains  a bag-of-words
because the ordering of the terms in the document is not modelled.

\subsection{Document Generation as a P\'olya process}

We use the P\'olya urn, and therefore the DCM, as a model for document
generation  where  the author  generates  an  actual  document $d$  by
drawing $|d|$  terms from the reinforced  document model. Intuitively,
different  documents  are written  in  different  styles (some  styles
exhibiting  more  word burstiness  than  others),  and therefore,  the
degree of  reinforcement will  be document specific.  Consequently, we
assume that each document is  drawn from a different document DCM, and
therefore we need  to estimate the parameters of  a different document
DCM for each document $d$.

The probability density function for the DCM is as follows:
% = \mathop{\mathbb{E_{\bm{\theta}|\bm{\alpha}}}}[p(d|\bm{\theta})]
\begin{equation} \label{eq:dcm}
p(d| \bm{\alpha}) = \int_{\bm{\theta}} p(d|\bm{\theta}) p(\bm{\theta}| \bm{\alpha}) d\bm{\theta} 
\end{equation}
where $\bm{\alpha}$ is  the initial $|v|$-dimensional parameter vector
of a Dirichlet distribution. Conceptually,  one can think of drawing a
multinomial $\bm{\theta}$  from a Dirichlet  distribution specified by
$\bm{\alpha}$,  and  subsequently  drawing   a  sample  $d$  from  the
multinomial.  The parameters  of the  DCM  can be  interpreted as  the
initial  number of  instances of  each  coloured ball  in the  P\'olya
urn. Therefore, the sum of the DCM parameter vector $\sum_{_{t \in v}}
{\alpha_t}$ can be  interpreted as the initial number  of balls in the
urn  (i.e.  $m_d   =  \sum_{_{t  \in  v}}  {\alpha_t}$)   and  is  the
concentration parameter.  This is the factor  that controls burstiness
on a  document level, and when  $m_d$ is large the  model exhibits low
burstiness as  adding balls to  the urn changes  the state of  the urn
very little.  In fact,  when $m_d  \to \infty$, the  DCM tends  to the
multinomial       distribution        (i.e.       no       burstiness)
\cite{elkan06}. Conversely,  if there  are very few  balls in  the urn
initially (i.e.  $m_d \to 0$),  the model exhibits high  burstiness as
the  first  ball  drawn  alters  the  initial  state  of  the  urn  by
reinforcement  quite  substantially. Therefore,  the  problem lies  in
estimating the initial parameters  of the document DCM $\bm{\alpha_d}$
given that  the document $d$  was generated by this  reinforced random
process.  For  consistency,  the   notation  we  use  to  specify  the
$|v|$-dimensional parameter  vector of the  DCM is similar to  that of
the Dirichlet distribution (i.e.  using a multinomial distribution and
a concentration parameter).

Furthermore, given that  documents only contain a subset  of the terms
in  the collection, we  do not  wish to  assign zero  probabilities to
terms  that do  not occur  in a  document. Therefore,  we  smooth each
document   DCM  ${\bm{\alpha}_{d}}$  with   a  background   DCM  model
${\bm{\alpha}_{c}}$.  The background  model is  the single  model most
likely  to have  generated \emph{all}  documents given  our reinforced
process, and therefore, we estimate the parameters of a background DCM
${\bm{\alpha}_{c}}$,  given  all  of  the  $n$  documents.  There  are
different ways in which we can smooth these two DCM models and we will
outline these  in Section~4.6.  In general, we  are not  restricted to
smoothing only two DCM models to construct our document model, and any
number  of plausible  DCM models  could  be combined  to help  explain
observations  in the  document. However,  in this  article  we confine
ourselves to smoothing only two DCM models for each document $d$.

\subsection{Non-Reinforced Query Likelihood}

Once  the   parameters  of   the  document  model   ($\mathcal{M}_d  =
\bm{\alpha}_{dm}$) have been estimated, we need to rank these document
models with  respect to  a query. In  the multinomial  language model,
both the document and query  are assumed, for the purposes of ranking,
to  have  been  generated  from  a multinomial.  This  simplifies  the
estimation  of the  document model  and  the estimation  of the  query
likelihood given the document model.

As  mentioned  earlier,  we  assume  that documents  and  queries  are
generated differently. More specifically we assume that queries do not
exhibit word burstiness. This  in fact simplifies the query likelihood
given  our  new document  model.  We  assume  that the  documents  are
generated from  a DCM document model $\bm{\alpha}_{dm}$,  and that the
query is generated  from the document model (urn)  using sampling with
replacement  (no reinforcement).  Modelling query  generation  in this
manner   means   that   each    term   in   the   query   is   treated
independently.  Consequently,   documents  are  ranked   according  to
following query likelihood formula:

\begin{equation} \label{eq:ql}
log \;  p(q|\mathcal{M}_d = \mathop{\mathbb{E}} [\bm{\theta}_{dm} | \bm{\alpha}_{dm}]) = log \; \prod_{t \in q} p(t | \mathcal{M}_d)^{c(t,q)} = \sum_{t \in q} (log \; p(t | \mathcal{M}_d) \cdot {c(t,q)})
\end{equation}
where  $\mathop{\mathbb{E}} [\bm{\theta}_{dm} |  \bm{\alpha}_{dm}]$ is
the expected multinomial of the DCM document model for document $d$.

\subsection{Estimation of the Document DCM}

\begin{figure}[ht] 
  \begin{center}
  	\begin{tabular}{c}
    \includegraphics[height=7cm,width=10cm]{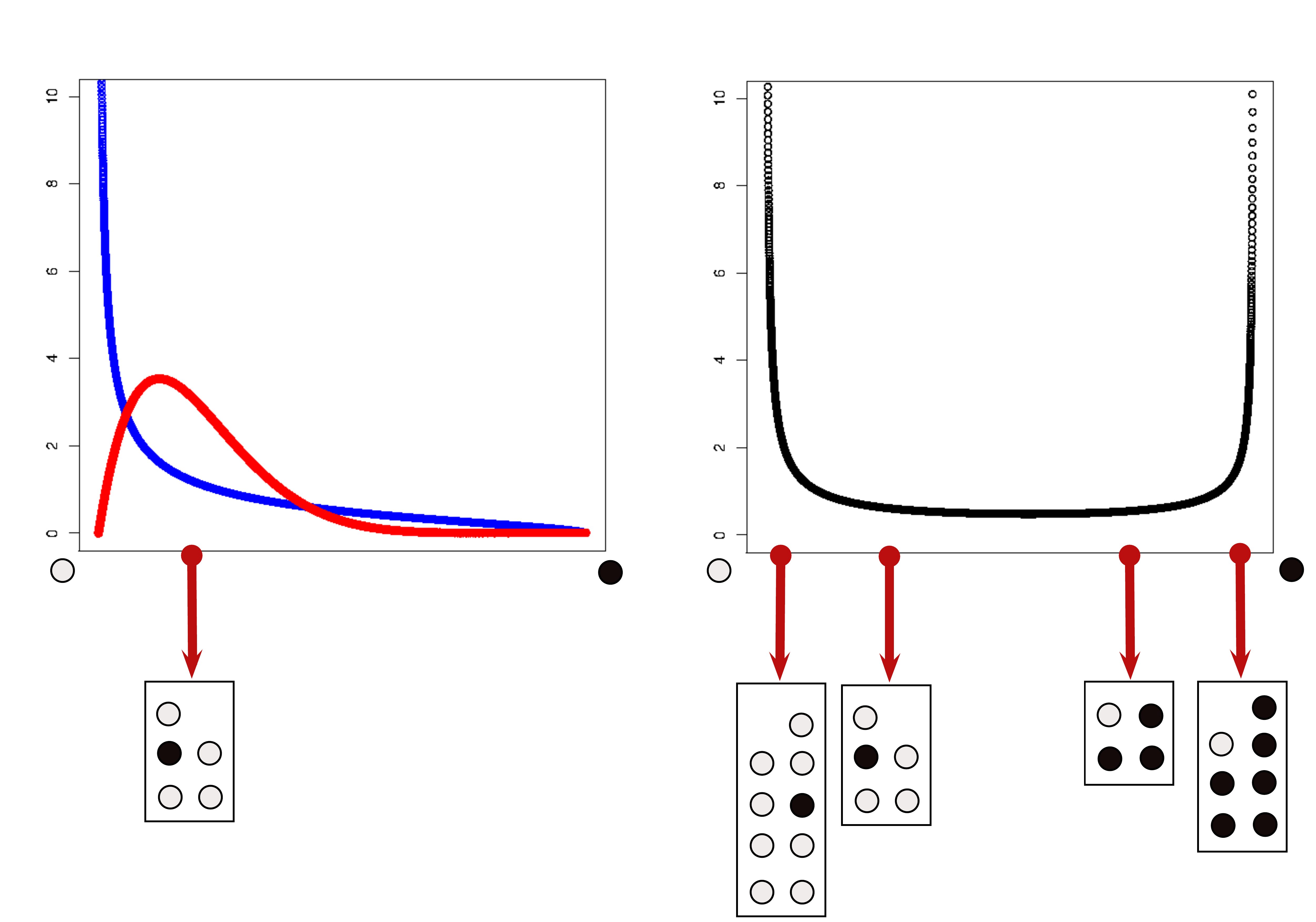}
     \end{tabular}
  \end{center}
  \caption{Documents   generated  from   multinomials  drawn   from  a
    Dirichlet  distribution for  both document  (left)  and background
    language models (right)}
  \label{figure:estimation}
\end{figure}

We  now estimate the  parameters of  the document  DCM $\bm{\alpha}_d$
using the  observations from the  actual document $d$. Given  only one
sample (i.e.  the document)  it is not  possible to fully  specify the
maximum likelihood estimates  of the document DCM\footnote{The minimum
  number  of samples  needed to  estimate both  the expected  value (a
  multinomial)  and   the  concentration  parameter   (burstiness)  is
  two.}. The maximum likelihood  estimates of the multinomial inferred
from one document will be equal to the expected value of the estimated
DCM. Therefore,  the maximum  likelihood estimates of  the multinomial
from which  the terms in  the document were drawn  (i.e. $c(t,d)/|d|$)
will  be  proportional to  the  maximum  likelihood  estimates of  the
document       DCM       (i.e.      $\bm{\hat{\theta}}_d       \propto
\bm{\hat{\alpha}}_d$). This  is only true  in the case where  there is
one sample.

Fig.~\ref{figure:estimation}  (left)  shows  this  graphically  for  a
simplified two-dimensional  model that uses  white and black  balls to
represent  terms.  The   x-axis  represents  multinomials  of  varying
parameter values. Points on the left-hand side of the x-axis represent
multinomials where the  probability of drawing a white  ball are high,
while  points   on  the  right-hand  side  of   the  x-axis  represent
multinomials  where  the  probability  of  drawing a  black  ball  are
high. The Dirichlet distribution  represents the likelihood of drawing
these   multinomials.  In   Fig~\ref{figure:estimation}   (left),  the
expectation  of both of  the two-dimensional  Dirichlet distributions,
shown  by  the  red and  blue  curves,  are  equal and  represent  the
multinomial (red arrow) inferred from the document.

Therefore,  when  we  have  only  one  multinomial  (inferred  from  a
document), we can only specify the location (expected multinomial) and
not  the shape  (concentration  parameter)  of the  DCM.  In order  to
completely define the parameters of  the document DCM, we also have to
define  the concentration  $m_d =  \sum_{_{t \in  v}} {\alpha_{d_t}}$,
which can  be interpreted as the  level of belief  associated with the
maximum likelihood  estimates of the  expected multinomial. Therefore,
the  initial  parameters of  the  $|{v}|$-component  document DCM  are
estimated as follows:

\begin{equation} \label{eq:doc_dcm}
\bm{\hat{{\alpha}}}_d = m_d \cdot \bm{\hat{\theta}}_{d} = ( m_d \cdot p(t_1|d), m_d \cdot p(t_2|d), .... , m_d \cdot p(t_{|v|}|d) ) 
\end{equation}
where $p(t|d) =  c(t,d)/|d|$ for all $t \in v$ and  where $m_d$ is the
initial   mass  that   controls   the  burstiness   of  the   document
model. Although estimation of the parameters of the DCM using multiple
data vectors  is computationally expensive \cite{minka00},  we can see
that estimating  the parameters of each  document DCM is  trivial if a
suitable value for $m_d$ can be found.

Given that $m_d$  is the level of belief  associated with the expected
document multinomial $\bm{\hat{\alpha}}_d$, it would seem intuitive to
aim to  minimise this  belief in the  absence of evidence  (an Occam's
razor  type  argument).  A  minimum   setting  can  be  arrived  at  by
determining the minimum initial number  of balls in the urn that could
have generated the document. Given  a document $d$, the minimum number
of balls initially in the urn is the number of distinct coloured balls
drawn. Therefore, we estimate the concentration parameter $m_d$ of the
document DCM as $\hat{m}_d = |\vec{d}|$. This is the maximum amount of
burstiness    that   is    supported   using    this    argument.   In
Fig.~\ref{figure:estimation}  (left)  our estimate  of  $m_d$ for  the
document model is  $m_d=2$, which leads to the  shape of the Dirichlet
in  blue.  Setting  $m_d$  according to  this  parsimonious  principle
ensures that we have not over-fitted to our data.

\subsection{Estimation of the Background DCM}

For  the  DCM  document  models,  there  exists  dependencies  between
successive occurrences of the same  term in a document, and therefore,
the  estimation of the  background DCM  is more  complex than  for the
multinomial distribution. In fact,  in the entire collection, the only
occurrences of  the same term that  are independent of  each other are
those  in different  documents. This  leads to  the introduction  of a
document boundary into  the background DCM of the  new language model,
something that is lacking in the multinomial language model.

The estimation of a background  DCM using all $n$ document vectors is,
as   mentioned   previously,   computationally   expensive.   However,
\citet{elkan06}  has   shown  that,  for  textual   data,  very  close
approximations to the maximum likelihood estimates of the DCM (via the
EDCM) are proportional  to $\sum_{j=1}^n I(c(t,d_j) > 0  )$ for all $t
\in v$, where $I$ is  the indicator function. These approximations are
accurate for textual  data because most terms do not  occur in all $n$
documents, and furthermore, it  has been shown that the approximations
make  little  difference  to   the  effectiveness  of  the  model  for
text-related tasks. It can be  seen that this approximation relates to
the  number of documents  in which  a term  occurs (i.e.  the document
frequency\footnote{This  introduces an  $idf$-like  measure into  this
  language model and is discussed in a later in Section~5.2.4}). Using
an appropriate  normalisation factor we obtain  a probability estimate
as follows:

\begin{equation} \label{eq:back_prob}
{p}(t|\bm{\hat{\theta'}}_c) = \frac{\sum_{j=1}^n I(c(t,d_j) > 0 ) }{\sum_{t' \in v} df_{t'}} = \frac{df_t}{\sum_{t' \in v} df_{t'}} = \frac{df_t}{\sum_{j=1}^n |\vec{d_j}|}
\end{equation}
where  $n$  is the  number  of documents  in  the  collection and  the
numerator  is the  document  frequency of  a  term. The  normalisation
factor can be  re-written and comprises the summation  of all document
vectors    in   the    collection    so   that    
$\sum_{t   \in    v} {p}(t|\bm{\hat{\theta'}}_c) = 1$.

This  probability   distribution  can   be  viewed  as   the  expected
multinomial  drawn from  the  background EDCM.  The  estimates of  the
background  DCM,   which  are  approximately   proportional  to  these
probability estimates, are defined in a similar manner to the document
DCM by introducing one  concentration parameter $m_c$. This results in
the following parameter estimates for the background DCM:

\begin{equation} \label{eq:back_dcm}
\bm{\hat{{\alpha}}}_c = ( m_c \cdot {p}(t_1|\bm{\hat{\theta'}}_c),  m_c \cdot {p}(t_2|\bm{\hat{\theta'}}_c), .... ,  m_c \cdot {p}(t_{|v|}|\bm{\hat{\theta'}}_c) )
\end{equation}
where  $m_c$ is  the belief  in the  expected value  of  the Dirichlet
(i.e. ${p}(t|\bm{\hat{\theta'}}_c)$) and can  be interpreted as a type
of document \emph{word burstiness} throughout the collection.

Fig.~\ref{figure:estimation}  (right) shows a  graphical example  of a
two-dimensional background Dirichlet. As before, the x-axis determines
parameter values  of the multinomials,  and the black curve  shows the
likelihood of  drawing these  multinomials. The four  document samples
shown in the figure exhibit  high levels of burstiness as they contain
a disproportionate  number of  balls of one  specific colour.  This is
because  areas of  higher  likelihood in  Fig.~\ref{figure:estimation}
lead  to multinomials  with one  component that  contains most  of the
probability mass. The convex shape of this curve is due to a low $m_c$
concentration  parameter, and  therefore  models high  levels of  word
burstiness.   Although   the   expectation  of   this   over-dispersed
two-dimensional  Dirichlet has  a  low likelihood,  it is  nonetheless
\emph{expected} in  the statistical sense.  Essentially, the use  of a
DCM explains greater term-frequency  variation in the $n$ documents in
the collection.

\subsection{Smoothing and Retrieval Models}

We now  present two  smoothing methods which  can be used  to linearly
combine $K$ multiple DCM models.

% Dirichlets
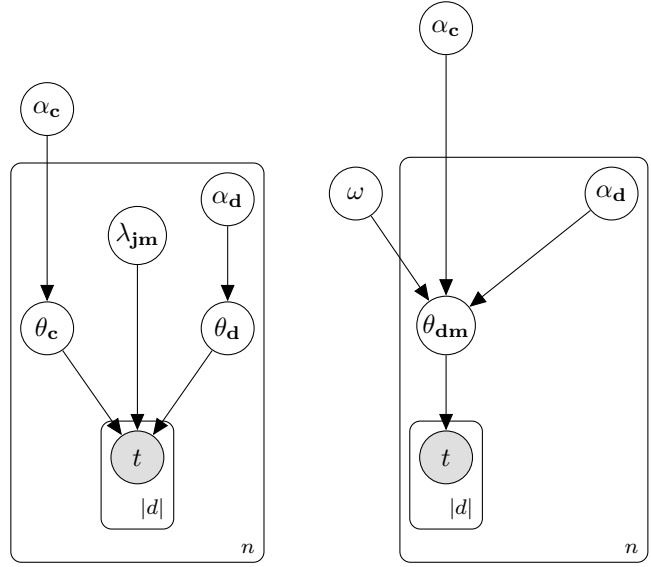
\begin{figure}[ht] 
  \begin{center}
  	\begin{tabular}{cc}
    % model_pca.tex
%
% Copyright (C) 2012 Jaakko Luttinen
%
% This file may be distributed and/or modified
%
% 1. under the LaTeX Project Public License and/or
% 2. under the GNU General Public License.
%
% See the files LICENSE_LPPL and LICENSE_GPL for more details.

% PCA model

%\beginpgfgraphicnamed{model-pca}
\begin{tikzpicture}

  % Define nodes
  \node[obs]                               		(term) 		{$t$};
  \node[latent, above=of term, xshift=1.2cm]  	(mult_d) 	{$\mathbf{\theta_d}$};
  \node[latent, above=of term, xshift=-1.2cm] 	(mult_c) 	{$\mathbf{\theta_c}$};
  \node[latent, above=of mult_c, xshift=0cm, yshift=1.2cm]   	(dir_c) 		{$\mathbf{{\alpha}_c}$};
  \node[latent, above=of mult_d, xshift=0cm]   	(dir_d) 		{$\mathbf{{\alpha}_d}$};
  \node[latent, above=of term, xshift=0.0cm, yshift=1.2cm]  (lambda) 	{$\mathbf{{\lambda_{jm}}}$};

  % Connect the nodes
  \edge {mult_d, mult_c, lambda} {term} ; %
  \edge {dir_d} {mult_d} ; %
  \edge {dir_c} {mult_c} ; %

  % Plates
  \plate {document} {(term) } {$|d|$} ;
  \plate {collection} {(term) (mult_d) (mult_c) (dir_d) (document.south west)} {$n$} ;

\end{tikzpicture}
%\endpgfgraphicnamed

%%% Local Variables: 
%%% mode: tex-pdf
%%% TeX-master: "example"
%%% End: 
	\hspace{2pt}		
	&
    \hspace{2pt}
    % model_pca.tex
%
% Copyright (C) 2012 Jaakko Luttinen
%
% This file may be distributed and/or modified
%
% 1. under the LaTeX Project Public License and/or
% 2. under the GNU General Public License.
%
% See the files LICENSE_LPPL and LICENSE_GPL for more details.

% PCA model

%\beginpgfgraphicnamed{model-pca}
\begin{tikzpicture}

  % Define nodes
  \node[obs]                               						(term) 		{$t$};
  \node[latent, above=of term]  									(mult_dm) 	{$\mathbf{\theta_{dm}}$};
%  \node[latent, above=of mult_dm]  									(dir_dm) 	{$\mathbf{\alpha_{dm}}$};  
  \node[latent, above=of mult_dm, xshift=0.0cm, yshift=2.2cm] 	(dir_c) 		{$\mathbf{{\alpha}_c}$};
  \node[latent, above=of mult_dm, xshift=2.2cm]   				(dir_d) 		{$\mathbf{{\alpha}_d}$};
  \node[latent, above=of mult_dm, xshift=-1.2cm]   				(omega) 		{$\mathbf{{\omega}}$};

  % Connect the nodes
  \edge {mult_dm} {term} ; %
%  \edge {dir_dm} {mult_dm} ; %
  \edge {dir_d} {mult_dm} ; %
  \edge {dir_c} {mult_dm} ; %
  \edge {omega} {mult_dm} ; %

  % Plates

  \plate {document} {(term) } {$|d|$} ;
  \plate {collection} {(term) (mult_dm) (dir_d) (document.south west) } {$n$} ;

\end{tikzpicture}
%\endpgfgraphicnamed

%%% Local Variables: 
%%% mode: tex-pdf
%%% TeX-master: "example"
%%% End: 
     \end{tabular}
  \end{center}
  \caption{Document generation in the SPUD model for both types of smoothing}
  \label{figure:plate}
\end{figure}

\subsubsection{Linear Smoothing of Expected Multinomials}

Conceptually, both the  background and document DCM can  be thought of
as a P\'olya urn. The first approach to smoothing treats each of these
models as  distinct P\'olya urns.  A document is generated  by drawing
with reinforcement,  balls from  the $K$ urns  according to  a certain
probability. Essentially this smoothing approach linearly combines the
expected  values  (multinomials)   of  the  Dirichlets.  This  general
smoothing approach is as follows:

\begin{equation} \label{eq:rbow_framework1}
p(d| \bm{\alpha}_{dm} ) = \sum^K_{i=1} \lambda_i \cdot p(d|{\bm{\alpha}}_i)
\end{equation}
where  $\sum^K_{i=1}  \lambda_i  =  1$ and  ${\bm{\alpha}}_i$  is  the
$i^{th}$ DCM model. In this  work we only linearly combine two models,
the  document DCM  and  the  background DCM,  and  therefore the  SPUD
retrieval model using this smoothing approach is defined as follows:

\begin{equation} \label{eq:rbow_framework2}
\mathop{\mathbb{E}}[ \bm{{\theta}}_{dm} | \bm{{\alpha}}_{dm} ] = (1 - \lambda_{jm}) \cdot \bm{\theta}_d  + \lambda_{jm} \cdot  \bm{{\theta}'}_{c} 
\end{equation}

where $\lambda_{jm}$ is the smoothing parameter and can be interpreted
as the  probability of  selecting a term  from the background  DCM. We
note    that   this    formulation   is    identical   to    that   of
\citet{hiemstra98}. Fig.~\ref{figure:plate} (left) shows the graphical
model for  the DCM language model  with this type  of linear smoothing
(Jelinek-Mercer).

One of  the main motivations for  smoothing the document  model with a
background model  is that the  background model assigns mass  to terms
unseen in  the document. Therefore, $\lambda_{jm}$  can be interpreted
as  the probability  of  drawing  a previously  unseen  term from  the
background model,  and $1-  \lambda$ as the  probability of  drawing a
previously seen term (i.e. a  repeated term from the document). During
the generation  of the document  $d$, at least  $|\vec{d}|$ previously
unseen   terms   were   drawn.   This   leads  to   an   estimate   of
$\hat{\lambda}_{jm} = |\vec{d}|/|d|$ as  the probability of drawing an
unseen  term  for that  document  model.  This  is the  proportion  of
distinct terms in the document and is the estimate of drawing from the
background  multinomial $\bm{{\theta'}_c}$.  The SPUD  retrieval model
with this type of smoothing is  denoted SPUD$_{jm}$ and it has no free
parameters. We note  that the estimation of ${\lambda}_{jm}$  is not a
consequence  of the DCM  model, and  can therefore  be applied  to the
multinomial language model that uses Jelinek-Mercer smoothing.

% Multinomials
%\begin{figure}[ht] 
%  \begin{center}
%  	\begin{tabular}{cc}
%	\input{./models/model_jm}  
%	\hspace{2pt}		
%	&
%    \hspace{2pt}
%    \input{./models/model_rbow_jm}
%     \end{tabular}
%  \end{center}
%  \caption{Original query likelihood model with linear smoothing and reinforced bow model with multinomial mixture}
%  \label{figure:plate_linear}
%\end{figure}

\subsubsection{Linear Smoothing of DCMs}

The  second approach to  smoothing uses  a linear  mixture of  the DCM
models. Conceptually, this approach to smoothing combines the contents
of  the $K$  urns into  one  single P\'olya  urn. A  document is  then
generated  by drawing with  reinforcement from  this single  urn. This
smoothing approach  is a more complete Bayesian  approach to smoothing
and the parameters of the document model are as follows:

\begin{equation} \label{eq:rbow_mix1}
\bm{{\alpha}}_{dm} = \sum^K_{i=1}{ \omega_i \cdot \bm{{\alpha}}_i }
\end{equation}
where  $\sum^K_{i=1}  \omega_i  =  1$  and  $\bm{{\alpha}}_i$  is  the
$i^{th}$ DCM language model. The $\omega$ parameters are linear mixing
parameters  that determine  the relative  weight of  the  DCM language
models. It is  worth noting that each of the  DCMs has a concentration
parameter $m_i$  which act to  weight the vector  appropriately. Given
the  document  DCM  and   background  DCM  estimated  previously,  the
smoothing is as follows:

\begin{equation} \label{eq:rbow_mix2}
\bm{\hat{\alpha}}_{dm} = (1 - \omega) \cdot m_d \cdot \bm{\hat{\theta}}_d + \omega \cdot m_c \cdot \bm{\hat{\theta'}}_c
\end{equation}
where $\omega$ is the linear mixing parameter. Fig.~\ref{figure:plate}
(right)  shows the  graphical model  for this  DCM mixture  model. The
expected  multinomial drawn  from  this DCM  mixture  model is  easily
computed using the individual parameters of the DCM mixture model over
the  normalisation  constant.  This  DCM mixture  retrieval  model  is
denoted SPUD$_{dir}$ due to the  mixing of the Dirichlets. Although it
seems  that the  DCM mixture  model still  has two  unknown parameters
(i.e. $m_c$  and $\omega$), these can  either be combined  to form one
single  parameter\footnote{This is analogous  to the  tuning parameter
  $\mu$  in  the  multinomial  language model  using  Dirichlet  prior
  smoothing.}, or  $m_c$ can be  estimated using numerical  methods as
outlined in the original  work introducing the EDCM \cite{elkan06}. We
outline the details of these approaches in the next section.

\section{Retrieval Model Implementation}

In  this section  we outline  the  composition of  the SPUD  retrieval
methods  using both  types  of smoothing  presented  in the  preceding
section.  We  then  present  some  retrieval intuitions  that  aid  in
understanding the retrieval aspects of the new model.

\subsection{Retrieval Functions}

Similarly  to the  implementation of  the standard  multinomial models
\cite{zhai04},  our  approach  can  be computed  efficiently  using  a
summation  that  only  involves  terms  common to  both  document  and
query. The SPUD$_{jm}$ retrieval function using linear smoothing is as
follows:

\begin{equation} \label{eq:rbow_jm1}
\bm{\textrm{SPUD}}_{jm}(q,d) =  \sum_{t \in q } (log ((1-\hat{\lambda}_{jm}) \cdot \frac{c(t,d)}{|d|} + \hat{\lambda}_{jm} \cdot \frac{df_t}{\sum_j^n |\vec{d_j}|}) \cdot c(t,q))
\end{equation}
where $\hat{\lambda}_{jm} = |\vec{d}|/|d|$. This is rank equivalent to the following:

\begin{equation} \label{eq:rbow_jm2}
\bm{\textrm{SPUD}}_{jm}(q,d) =  |q|\cdot log(\hat{\lambda}_{jm}) + \sum_{t \in q \cap d} (log (1 + \frac{(1-\hat{\lambda}_{jm}) \cdot {c(t,d)} \cdot {\sum_j^n |\vec{d_j}|}}{ {|\vec{d}|}  \cdot {df_t}  }) \cdot c(t,q))
\end{equation}
The  SPUD$_{dir}$ retrieval  function can  be computed  in  a somewhat
similar form  to the multinomial language model  using Dirichlet prior
smoothing as follows:

\begin{equation} \label{eq:rbow_dir1}
\bm{\textrm{SPUD}}_{dir}(q,d) =  \sum_{t \in q } (log ( \frac{ (1- \omega) \cdot |\vec{d}| \cdot \frac{c(t,d)}{|d|} +  {\omega \cdot m_c} \cdot \frac{df_t}{\sum_j^n |\vec{d_j}|}}{(1-\omega)\cdot |\vec{d}| + \omega \cdot m_c} ) \cdot c(t,q))
\end{equation}
which is rank equivalent to the following:
\begin{equation} \label{eq:rbow_dir2}
\bm{\textrm{SPUD}}_{dir}(q,d) = |q|\cdot log(\frac{\mu'}{\mu' + |\vec{d}|}) + \sum_{t \in q \cap d} (log (1 + \frac{ |\vec{d}| \cdot c(t,d) \cdot {\sum_j^n |\vec{d_j}|}}{\mu' \cdot |d| \cdot df_t}) \cdot c(t,q) )
\end{equation}
where $\mu'$ is a combination of $\omega$ and $m_c$ as follows:

\begin{equation} \label{eq:rbow_mixing_param}
\mu' = \frac{\omega}{1 - \omega} \cdot m_c
\end{equation}
As $\mu'$ is the only parameter that has not been estimated so far, we
now  outline two  approaches to  finding suitable  values for  it. The
first approach is to experimentally  tune $\mu'$ on training data in a
similar manner to the Dirichlet prior smoothing parameter $\mu$ in the
multinomial language model  \cite{zhai01}. Alternatively, $m_c$ can be
estimated from  the $n$ samples of observations  using Newton's method
\cite{elkan06} as follows:

\begin{equation} \label{eq:estimate_w}
m_{c}^{new} = \frac{\sum_j^n |\vec{d_j}|} {\sum_j^n \psi(|d_j| + m_c) - n \cdot  \psi(m_c)} 
\end{equation}
where $\psi(x)  = \frac{d}{dx}log  \Gamma(x)$ is the  digamma function
and $\Gamma$  is the  gamma function. When  estimating $m_c$  from the
data  using  this method,  $\omega$  is  the  parameter that  requires
experimental tuning. However,  we expect that the one  setting for the
hyperparameter  $\omega$  will   perform  robustly  across  many  test
collections. Experiments for both of these approaches to determining a
suitable values the free parameters are are outlined in Section~6.4.

\subsection{Length Normalisation and Document Boundary Retrieval Intuitions}

We now examine some  retrieval intuitions and existing hypotheses that
help explain the differences  between the SPUD retrieval functions and
the multinomial retrieval functions. For  most of the analysis in this
section   we  focus   on   the  best   performing  multinomial   model
(MQL$_{dir}$)    and   its   counterpart    from   the    SPUD   model
(SPUD$_{dir}$).   \citet{robertson1994}    outlined   two   hypotheses
concerning  the  length  of  a document,  namely  the  \emph{verbosity
  hypothesis} and the \emph{scope hypothesis}, which we now examine.

\subsubsection{Verbosity Hypothesis}

The  \emph{verbosity  hypothesis}  captures  the intuition  that  some
documents  are  longer  than  others  simply  because  they  are  more
verbose. Such documents  do not describe more topics,  they are simply
more  \emph{wordy}. This  hypothesis  captures an  aspect of  document
length  that  is  independent   of  relevance.  However,  the  initial
description of  this hypothesis \cite{robertson1994}  does not outline
any  formal  means  of  determining  whether  a  particular  retrieval
function is consistent with the hypothesis. We now outline a retrieval
constraint\footnote{Just prior to publication  we found that a similar
  constraint  has  been previously  been  outlined in  \cite{na2008}.}
which helps to determine this.

\newtheorem*{axiom}{LNC2*}
\begin{axiom}
If document $d$ and $d'$  are two documents, where $d'$ is constructed
by  concatenating  $d$ with  itself  $k$  times  where $k>0$,  and  if
$s(q,d)$ is the score returned  from a retrieval function $s$ which is
used to rank $d$ with respect to $q$, then $s(q,d) = s(q,d')$.
\label{LNC2*}
\end{axiom}
This states that if a  document is concatenated with itself any number
of times, the retrieval score of that document should not change for a
given query,  and therefore  it should not  change rank. We  call this
constraint  {LNC2*}  as  this  is  stricter than  {LNC2}  outlined  by
\citet{fang04},    which    only     states    that    $s(q,d)    \leq
s(q,d')$. Essentially  if a scoring  function $s$ adheres  to {LNC2*},
then we deem $s$ to be consistent with the verbosity hypothesis.

Consider a relevant document $d$  that is ranked in a certain position
according to $s(q,d)$. If $d$ is replaced in the collection with $d'$,
$d'$   should  not  be   ranked  lower   than  the   initial  document
$d$. Therefore,  $s(q,d')$ should certainly not be  less than $s(q,d)$
simply  due to  the verbosity  of  $d'$. Now  consider a  non-relevant
document $d$ of  a given length. If $d$ is  replaced in the collection
with $d'$,  $d'$ should not  be ranked in  a higher position  than $d$
originally was. Therefore, given that  we do not know the relevance of
$d$ a priori,  we argue that in general  $s(q,d')$ should not increase
simply due to the increased verbosity of $d'$.

The   maximum  likelihood   estimate   of  a   term   in  a   document
(i.e. $c(t,d)/|d|$)  will not change if that  document is concatenated
with itself any number of  times. However, in the multinomial language
model using Dirichlet priors  smoothing (MQL$_{dir}$), {LNC2*} is only
satisfied  when  $c(t,d)/|d|  =  cf_t/|c|$  which  is  not  often  the
case. For this model, if there are many query-term matches in $d$, the
more verbose  document $d'$ will  nearly always be ranked  higher than
$d$ (i.e.  $s(q,d') > s(q,d)$)\footnote{We  note that we  are ignoring
  the effect  that creating a longer  document $d'$ would  have on the
  background collection model. For  an extremely large collection this
  effect  would   be  negligible.   Furthermore,  we  note   that  the
  multinomial  language  model   that  uses  Jelinek-Mercer  smoothing
  adheres to LNC2*, while the SPUD$_{jm}$ does not. However, there are
  other reasons  for the generally weaker performance  of the standard
  multinomial language model with Jelinek-Mercer smoothing.}, while if
there are very few query-term matches in $d$ the verbose document $d'$
will  nearly  always  be  ranked  lower  than  $d$  (i.e.  $s(q,d')  <
s(q,d)$).  However,   if  we   examine  the  SPUD$_{dir}$   method  in
Eq.~(\ref{eq:rbow_dir2}), we  can see that the  document vector length
$|\vec{d}|$ is used as one  form of document length normalisation. The
document  vector  length $|\vec{d}|$  will  remain  unchanged for  the
concatenated        document         $d'$,        and        therefore
$\bm{\textrm{SPUD}}_{dir}(q,d) = \bm{\textrm{SPUD}}_{dir}(q,d')$.

\begin{figure}[!ht] 
\begin{center}
$\begin{array}{c c}
	\includegraphics[height=4cm,width=5cm]{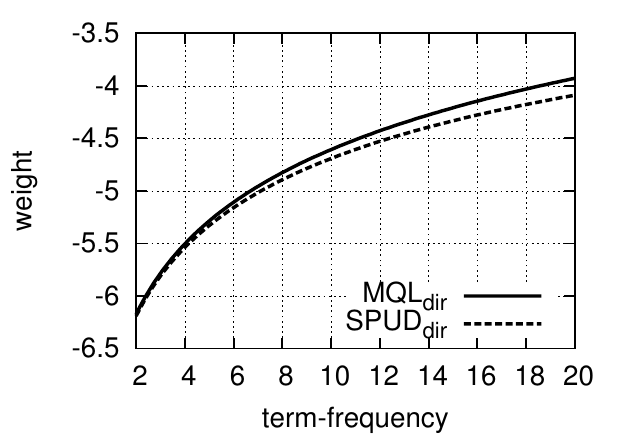} &
	\includegraphics[height=4cm,width=5cm]{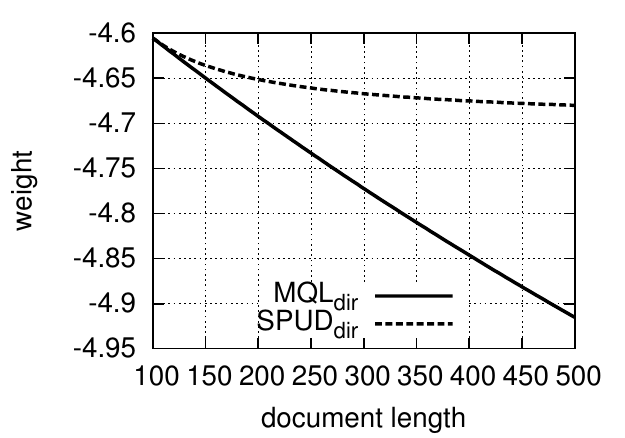}
\end{array}$
\caption{Change in weight  as term-frequency increases for MQL$_{dir}$
  and SPUD$_{dir}$ in a  document that initially contains 100 distinct
  terms  (left). Change  in weight  as recurring  non-query  terms are
  added  to a  document  that initially  contains  100 distinct  terms
  (right).}
\label{fig:tf}
\end{center}
\end{figure}

In general the multinomial model not only over-promotes recurrences of
query  terms but over-penalises  recurrences of  non-query terms  in a
given document. Fig.~\ref{fig:tf} (left)  shows the increase in weight
as   the   term-frequency   increases   for   both   MQL$_{dir}$   and
SPUD$_{dir}$. We  can see that  MQL$_{dir}$ gives a greater  weight to
terms with  higher frequencies than SPUD$_{dir}$. This  is because the
aspect of document length that is affected as term-frequency increases
is  different  for both  retrieval  functions.  It  is important  that
term-frequency is  analysed considering the change  in document length
that  an increase  in term-frequency  brings  about. Fig.~\ref{fig:tf}
(right) also  shows the penalisation  due to recurrences  of non-query
terms  for  both  MQL$_{dir}$   and  SPUD$_{dir}$.  We  can  see  that
MQL$_{dir}$  penalises  recurrences   of  non-query  terms  more  than
SPUD$_{dir}$. In  the SPUD$_{dir}$  function, recurrences of  the same
non-query term  will always  decrease the score  of a document  due to
more off-topic verbosity.

Interestingly,  it  can be  seen  that the  $\bm{\textrm{SPUD}}_{dir}$
formula  in Eq.~(\ref{eq:rbow_dir2})  contains the  ratio  between the
term-frequency $c(t,d)$ and the average term-frequency in the document
$|d|/|\vec{d}|$.  This average  term-frequency normalisation  idea was
first proposed by  \citet{singhal96}, but in general was  not shown to
improve  retrieval effectiveness  substantially until  recent research
\cite{paik:13}.  It  is this  part  of the  $\bm{\textrm{SPUD}}_{dir}$
retrieval model that deals specifically with the verbosity hypothesis,
while the document length  normalization component, the left-hand side
of  Eq.~(\ref{eq:rbow_dir2}),  deals  with  the  scope  hypothesis  by
replacing  the  original  document  length with  the  document  vector
length. We now discuss this further.

\subsubsection{Scope Hypothesis}

The  \emph{scope hypothesis} captures  the alternative  intuition that
documents may be  longer because they cover many  different topics. It
has been  noted in  the original work  regarding the  scope hypothesis
\cite{robertson1994} that many Newswire documents in the original TREC
corpora  seemed  as  if  they  consisted of  multiple  different  news
articles  concatenated  together. In  the  multinomial language  model
there is no difference in the normalisation applied when a term occurs
for the  first time (i.e. an increase  in scope) as opposed  to when a
term repeats  itself (i.e. an increase in  verbosity). This difference
is modelled in the new SPUD language models and can be viewed as being
modelled separately for  SPUD$_{dir}$. In Eq.~(\ref{eq:rbow_dir2}), we
can see that the factor  $|q| \cdot log(u'/(u' + |\vec{d}|))$ leads to
a penalisation  only for the  occurrence of distinct terms  (i.e. when
the  scope broadens\footnote{We  assume  that the  number of  distinct
  terms in  a document  is a  crude measure of  scope.}). If  the term
re-occurs, it is  not penalised by the part  of the retrieval function
which deals with scope.

For the  SPUD$_{dir}$ model, adding  a non-query term into  a document
for  the first  time will  lead to  penalisation by  the normalisation
aspect that  deals with  scope. However, it  should be noted  that the
verbosity aspect  of a  document is also  affected by the  addition of
previously unseen non-query terms  and this actually promotes existing
query-terms.   Therefore,  the   overall  document   score   does  not
necessarily decrease when a new non-query term is added.

In  the  SPUD$_{dir}$  model,  the  magnitude of  the  document  score
penalisation for  the first  occurrence of a  non-query term  is quite
similar   to   the    penalisation   applied   by   MQL$_{dir}$   (See
Fig.~\ref{fig:tf}),  but  recurrences   are  \emph{not}  penalised  as
much. Given  these observations, we hypothesise  that the SPUD$_{dir}$
retrieval  method  does  not   penalise  long  documents  as  much  as
MQL$_{dir}$. Recent research has studied the over penalisation of long
documents   by   many   retrieval  functions   including   MQL$_{dir}$
\cite{lv11}. They  built upon  work by \citet{singhal96}  which showed
that most ranking functions  retrieve long documents with a likelihood
less than their likelihood of relevance. We replicate that analysis by
binning according  to length,  relevant documents and  then estimating
the probability that  a document occurs in a  given bin (length) given
that  it is  relevant.  The  same procedure  is  applied to  retrieved
documents where a document is deemed retrieved if it occurs in the top
1000 documents of the ranked list. We use the same binning strategy as
\citet{lv11} (i.e. 5000) and compared the MQL$_{dir}$ and SPUD$_{dir}$
retrieval functions.  The aspect  of length used  in this  analysis is
that of the number of word tokens in the document (i.e. $|d|$).

\begin{figure}[!ht] 
\begin{center}
$\begin{array}{c c c}
   	\includegraphics[height=4cm,width=5cm]{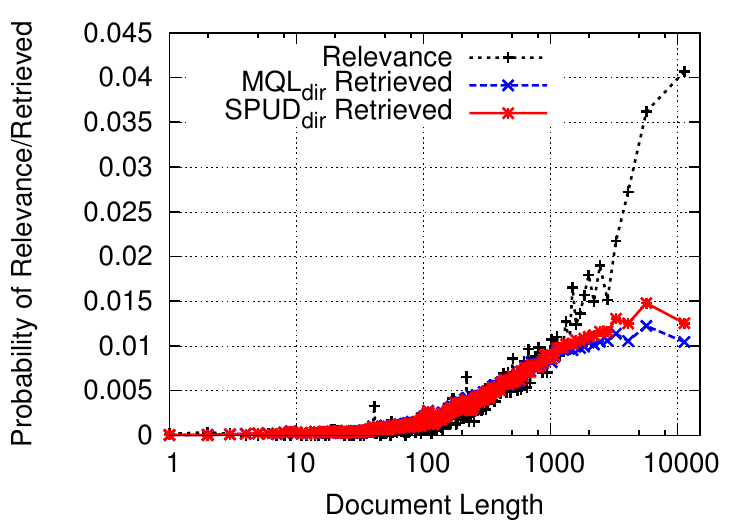} &    
	\includegraphics[height=4cm,width=5cm]{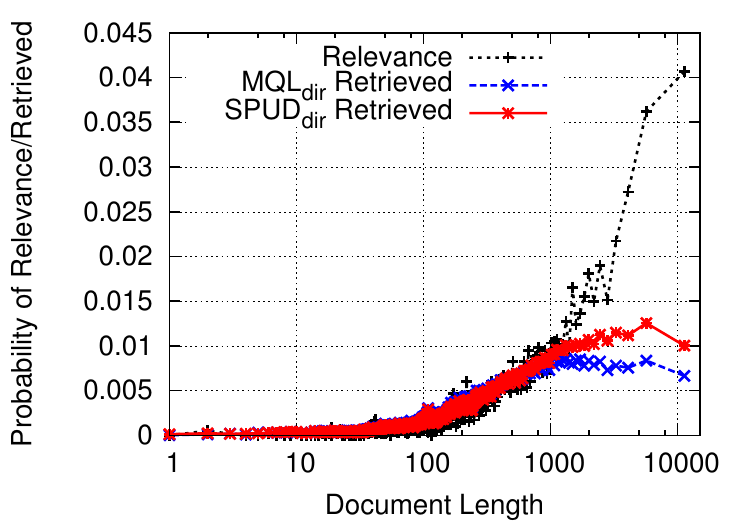} \\
\end{array}$
\caption{Probability   of  retrieval/relevance  for   MQL$_{dir}$  and
  SPUD$_{dir}$  methods  for trec-9/10  collection  for short  queries
  (left) and medium length queries (right). }
\label{fig:prob_rel}
\end{center}
\end{figure}

Fig.~\ref{fig:prob_rel} shows the  probability of relevance (in black)
and the  probability of retrieval  for both MQL$_{dir}$ (in  blue) and
SPUD$_{dir}$ (in  red) on  one collection. Firstly,  we note  that the
trends    are     consistent    with    the     previous    approaches
\cite{singhal96,lv11}. Furthermore,  we can see  that longer documents
have  a  higher likelihood  of  being  retrieved  by the  SPUD$_{dir}$
approach  compared  to the  MQL$_{dir}$  approach.  This confirms  our
intuitions  that  the  SPUD$_{dir}$   model  does  not  penalise  long
documents as  much as  MQL$_{dir}$ and that  we would expect  the SPUD
method to retrieve  long documents with a probability  closer to their
likelihood  of   relevance.  We   investigate  this  further   in  the
experimental section (Section 6.5).

\subsubsection{Background model}

% somewhere in your document's body:
\begin{table}[!ht] 
\tbl{Sample collection of four documents and two terms \label{tab:background_example}}{
\centering
\renewcommand{\arraystretch}{1.3}
\setlength\tabcolsep{10pt}
\begin{tabular}{|l|l|l|}

\hline
docs & $t_1$ & $t_2$ \\
\hline
\hline
$d_1$ & 8 & 2 \\
\hline
$d_2$ & 0 & 1 \\
\hline
$d_3$ & 0 & 3 \\
\hline
$d_4$ & 0 & 1 \\
\hline
\end{tabular}}
\end{table}

The  new  background  model  in  the  SPUD  brings  about  some  other
interesting retrieval characteristics.  Given the sample collection in
Table  \ref{tab:background_example} of  four documents  and  two terms
($t_1$ and $t_2$), we might wish to determine the most likely one term
string,  $q=\{t_1\}$  or $q=\{t_2\}$,  generated  from the  background
model.  If we assume  a multinomial  background model  estimated using
maximum  likelihood,  then  $p(t_1|\bm{\hat{\theta}}_c)  =  8/15$  and
$p(t_2|\bm{\hat{\theta}}_c) = 7/15$,  suggesting that term $q=\{t_1\}$
is  the  more  likely.  However,  intuitively we  see  that  the  high
frequency  of  ${t_1}$  in   document  $d_1$  is  unduly  biasing  the
estimates,   especially  as   term   ${t_1}$  only   appears  in   one
document. Term ${t_2}$ occurs in  all of the documents, and therefore,
is a word used more widely in the collection (possibly by more authors
in general). The  SPUD model takes the document  boundary into account
yielding   estimates   of   $p(t_1|\bm{\hat{\theta'}}_c)=   1/5$   and
$p(t_2|\bm{\hat{\theta'}}_c) = 4/5$  respectively. This probability is
similar  to that  proposed  in  one of  the  first language  modelling
approaches  \cite{hiemstra98}, and  has recently  been  re-examined as
being potentially theoretically valid \cite{roelleke12}.

\begin{figure}[!ht] 
\begin{center}
$\begin{array}{c c c}
	\includegraphics[height=4cm,width=5cm]{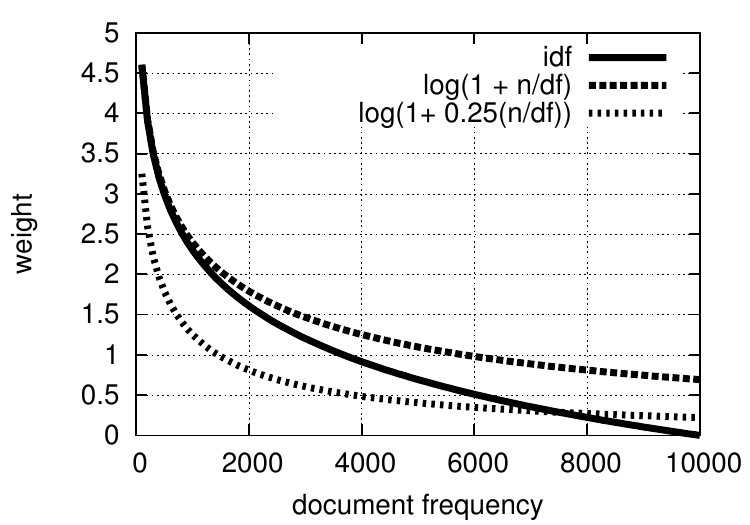} &
\end{array}$
\caption{idf and global weightings derived from the SPUD model}
\label{fig:idf}
\end{center}
\end{figure}

As  seen in this  toy example,  the proposed  model uses  the document
frequency in  its approximation for  the parameters of  the background
DCM. Furthermore,  the normalisation component used  in the background
model   can  be   written   as  $\sum_j^n   |\vec{d_j}|   =  n   \cdot
|\vec{d}|_{avg}$,  where  $|\vec{d}|_{avg}$  is the  average  document
vector length.  Therefore, in  the SPUD$_{dir}$ retrieval formula, the
weight assigned to a query-term that occurs in a document comprises of
the    following   factor    as   per    the   right-hand    side   of
Eq.~(\ref{eq:rbow_dir2}):
\begin{equation} \label{eq:rbow_idf}
\sum_{t \in q \cap d} log (1 + \delta \cdot \frac{n}{df_t}) \cdot c(t,q)
\end{equation}
where $\delta  = |\vec{d}| \cdot |\vec{d}|_{avg} \cdot  c(t,d) / (\mu'
\cdot |d|  )$. We  can see  that this factor  can be  viewed as  a new
family  of idf.  Unlike the  traditional idf  measure, this  factor is
document-length,     document-vector-length,     and    term-frequency
specific\footnote{We   note  that  \citet{sunehag07}   has  previously
  derived the  traditional idf from  a P\'olya process  using slightly
  different  assumptions.}.  We  have  found that  $\delta$  typically
ranges from $0.05$ to $0.5$ for query terms on many of the collections
used in this work. Fig.~\ref{fig:idf} shows the weight assigned by idf
and by Eq.~(\ref{eq:rbow_idf}) as the document frequency changes. This
suggests  that the  global weighting  factor  in our  new approach  is
closely related  to idf. This  crude comparison by no  means validates
the traditional idf in a theoretical perspective, nevertheless it does
present a  theoretical means by  which aspects of  both term-frequency
and  document  frequency  combine  in  one  model.  In  contrast,  the
multinomial  language   modelling  approach  treats   terms  that  are
completely independent of each  other, written by different authors on
different topics, similarly to terms that are highly dependent on each
other   (e.g.    terms   that   are   repeated,    possibly   due   to
\emph{association},  in  a  document   written  by  one  author  on  a
particular  topic). Discovering  a theoretical  justification  for the
combination  of both  term-frequency and  idf is  problematic  as they
appear    to    lie    in    different    event    spaces\footnote{See
  \cite{robertson04} for a thorough  review of theoretical attempts to
  justify   idf    with   term-frequency.}.   The   \emph{preferential
  attachment}  captured in the  SPUD model  is a  promising generative
theory justifying tf-idf type schemes.

A practical  consideration is whether there  is substantial difference
between  the  probability of  a  term  given  either background  model
(multinomial or DCM) when estimated from data. Therefore, we estimated
the background  probability of seeing a  term for both  models for all
query terms on one of the test collections used in our experiments. We
analysed 1530 query-terms from  the {trec-9/10} test collection and we
found  a  high  linear   correlation  (0.954)  between  the  estimated
probabilities  for  the   terms.  This  is  to  be   expected  as  the
probabilities are fundamentally  capturing similar information about a
term. However, there are examples where the estimated probabilities of
actual               query-terms               are               quite
different. Table~\ref{tab:background_terms}  shows the top  and bottom
10 terms  when ranked  according to the  ratio of  their probabilities
(i.e.  $p(t|\bm{\alpha}_c)/p(t|\bm{\theta}_c)$). The  bottom  10 terms
show  those  that  the  background  multinomial gives  a  much  higher
probability to when compared to  the background DCM. It is interesting
that  the term  \emph{el}, which  has much  higher probability  in the
multinomial  model, is  a  stopword from  a  different language.  This
receives  a relatively  high probability  estimate from  a multinomial
because it appears  many times, but receives a  much lower probability
estimate  from the background  DCM because  many of  these appearances
come  from  few  documents  (i.e.  the  term  is  quite  bursty).  The
background  DCM   regards  these  terms  as  less   general  than  the
multinomial,  as  the  occurrences  have actually  occurred  in  fewer
documents.  Conversely,   the  top  10  terms  show   those  that  the
multinomial  model  has  estimated  as  less  general  but  which  the
background DCM  has estimated as  being more general. These  terms are
less bursty but have occurred in many documents in the collection.

Therefore, given  that there exists query-terms  where the probability
of occurrence under our new  model is quite different, we would expect
this to  impact retrieval effectiveness.  We evaluate the  effect that
the new document normalisation  and background model have on retrieval
effectiveness separately in Section~6.5.

\begin{table}[!ht] 
\tbl{Ratio of estimated query-term probabilities of DCM to multinomial model ($p(t|\bm{\alpha}_c)/p(t|\bm{\theta}_c)$) for a number of query-term on {trec-9/10} \label{tab:background_terms}}{
\centering
\renewcommand{\arraystretch}{1.3}
\setlength\tabcolsep{10pt}
\begin{tabular}{|l||l|l||l|l|}

\hline
terms &  \multicolumn{2}{c|}{Bottom 10}  &  \multicolumn{2}{c|}{Top 10}\\
\hline
1	&	vike		&	0.3461802368		&	funnel-shap	&	2.0327764077		\\
2	&	el		&	0.3910938927		&	undergon		&	1.9616724275		\\
3	&	cancer	&	0.4098663257		&	pejor		&	1.9517391904		\\
4	&	patient	&	0.415028517		&	tartin		&	1.9495633384		\\
5	&	cell		&	0.4180174157		&	superstiti	&	1.9149743114		\\
6	&	student	&	0.4289654263		&	gynt			&	1.9071815267		\\
7	&	drug		&	0.4726560064		&	interest		&	1.8928123508		\\
8	&	system	&	0.4950172629		&	unsuccess	&	1.8803346212		\\
9	&	law		&	0.5064728539		&	work			&	1.8709780476		\\
10	&	infect	&	0.51433562		&	run-awai		&	1.8666031964		\\
\hline

\hline
\end{tabular}}
\end{table}

\section{Experiments}

In this  section we outline the  experiments used to  evaluate the new
SPUD   methods.  We   first  outline   the  experimental   design  and
methodology, before presenting the experiments.

\subsection{Experimental Design}

We carry out four experiments to evaluate different aspects of the new
SPUD  query  likelihood models.  The  first  experiment evaluates  the
retrieval effectiveness  of the new  SPUD retrieval methods  against a
number of baselines. The second experiment evaluates the robustness of
the  tuning  parameters  in  the  SPUD retrieval  methods.  The  third
experiment presents  an analysis of the  retrieval intuitions outlined
in the preceding section. Finally, we evaluate the best SPUD retrieval
method  when   incorporating  it  into   a  pseudo-relevance  feedback
framework.

\subsection{Datasets}

Table   \ref{tab:collections}  shows   the   characteristics  of   the
TREC\footnote{http://trec.nist.gov/}  test  collections  used  in  the
experiments. We  use a  wide variety of  TREC collections that  are of
varying  sizes  and include  collections  of  Web documents,  Newswire
articles, and medical abstracts.  In our experiments we evaluate short
keyword queries (2-3 terms) consisting  of the title field of the trec
topic, medium  queries (6-10 terms)  consisting of both the  title and
description  fields of  the topics,  and long  verbose  queries (10-30
terms) consisting  of the title, description, and  narrative fields of
the  topic. We  remove  standard stopwords  and  apply stemming  using
Porter's stemmer. It is worth  noting that the ohsumed test collection
contains only description length queries (i.e. medium length queries),
while there are only title  length queries available for the mq-07 and
mq-08 test collections.

\begin{table}[!ht]
%\scriptsize
\tbl{Test Collection Details  \label{tab:collections}}{
\centering
\renewcommand{\arraystretch}{1}
\setlength\tabcolsep{3pt}
\begin{tabular}{| l | l || r | c | c | c | c | c |}
\hline

	 \multicolumn{5}{| c |}{}  	& \multicolumn{3}{| c |}{query length}	\\

\hline
	label & \multicolumn{1}{| c ||}{collection}  	& \# docs &  \# topics &  topic range  	&  short 	& medium	& long \\
	  & \multicolumn{1}{| c ||}{ }  	&   &    &  				 &  (title)	& (title+desc)	& (title+desc+narr)\\	
\hline
%Tuning &	LATIMES	& 131,896		& 144		& 301-450	\\
 %	&	WSJ		& 130,837		& 150		& 051-200	\\
\hline	

ohsu			&	ohsumed			& 293,856 		& 63 		& 001-63 			& n/a	& 5.0		& n/a	\\
robust-04		&	fr, ft,la, fbis	& 528,155		& 250		& 301-450, 601-700	& 2.5	& 10.3		& 31.4	\\
trec-8			&	wt2g			& 221,066 		& 50		& 401-450			& 2.4	& 9.0		& 27.5	\\
trec-9/10		&	wt10g			& 1,692,096		& 100		& 451-550			& 2.6	& 9.3		& 24.3	\\
gov2			&	gov2			& 25,205,179	& 150		& 701-850			& 2.8	& 8.6		& 33.3	\\
mq-07			&	gov2			& 25,205,179	& 1778		& 1-10k				& 3.1		& n/a		& n/a 	\\
mq-08			&	gov2			& 25,205,179	& 784		& 10k - 20k			& 3.7		& n/a		& n/a 	\\

\hline
\end{tabular}}
\end{table}

\subsection{Retrieval Effectiveness}

The first experiment evaluates the retrieval effectiveness of the SPUD
model  against   its  counterpart,  the   standard  multinomial  query
likelihood  language  model.  We  compare  the  SPUD$_{jm}$  retrieval
function  against  the  multinomial  query  likelihood  function  with
Jelinek-Mercer smoothing (MQL$_{jm}$). We tune the MQL$_{jm}$ function
for each  set of queries to  optimise mean average  precision (MAP) on
each test  collection where the  parameter space $\pi_{jm}  \in \{0.1,
0.2,  ...,  0.9,  1.0\}  $.  Therefore,  we  are  confident  that  the
effectiveness  of the MQL$_{jm}$  retrieval function  is close  to its
optimal  on  each  collection. On  the  other  hand,  we do  not  tune
SPUD$_{jm}$ as it has no free-parameters.

We   compare   the  SPUD$_{dir}$   retrieval   function  against   its
counterpart,   the   multinomial   query  likelihood   function   with
Dirichlet-prior  smoothing  (MQL$_{dir}$).   Similarly,  we  tune  the
MQL$_{dir}$ function to optimise MAP on each test collection where the
parameter space  $ \mu \in \{250,  500, ..., 2250,  2500\}$. We report
the  effectiveness of  the  SPUD$_{dir}$ retrieval  function for  same
parameter setting as MQL$_{dir}$  (i.e. $\mu = \mu'$). This evaluation
favours MQL$_{dir}$ as SPUD$_{dir}$ may not be tuned optimally.

We also  use the  DCM-L-T retrieval function  \cite{xu08} which  has a
tuning parameter $\gamma$.  We tuned $\gamma$ for each  set of queries
on each  collection over the  parameter space $\gamma \in  \{0.1, 0.2,
...,  0.9,  1.0\}$\footnote{The  original  paper does  not  outline  a
  recommended parameter space. However when tuning from $0.1 - 1.0$, a
  maximum stationary point for effectiveness was found for each set of
  queries.}.

\subsubsection{Retrieval Effectiveness Results}

Tables  \ref{tab:results_title_map}  and  \ref{tab:results_title_ndcg}
show  the  retrieval effectiveness  (MAP  and  NDCG@20) of  MQL$_{jm}$
compared to SPUD$_{jm}$, and  MQL$_{dir}$ compared to SPUD$_{dir}$ for
short title queries (2-3 terms on average). We can see that on most of
the  test  collections  the   SPUD  retrieval  methods  demonstrate  a
significant increase  in effectiveness for  both MAP and  NDCG@20 over
their corresponding MQL methods.

\begin{table}[!ht] 
%\scriptsize

\tbl{MAP of SPUD models vs MQL models  ($\blacktriangle$ means two-sided t-test $p < 0.01$,  
$\triangle$ means $p < 0.05$) and SPUD models vs DCM-L-T ($\bullet$ means 
two-sided t-test $p  < 0.01$ compared to DCM-L-T, $\circ$ means $p    <   0.05$ 
compared to DCM-L-T). \label{tab:results_title_map}}{
\centering
\renewcommand{\arraystretch}{1.1}
\setlength\tabcolsep{3pt}
\begin{tabular}{| l || l | l | l | l | l | l | l | l | l |}

\hline	
				&		 \multicolumn{6}{c|}{short queries} 	\\
\hline
				&  	  {robust-04} 		& {trec-8} & {trec-9/10} & 	{gov2} &	{mq-07}	&	{mq-08}	\\

\hline
DCM-L-T			&		0.248	&	0.306	&	0.187	&	0.288	& 0.409 	& 0.413			 \\
%\hline
\hline
MQL$_{jm}$		&		0.231	&	0.246	&	0.135	&	0.245	& 0.396 	& 0.419			 \\
SPUD$_{jm}$		&		0.236	&	0.255	&	0.154$\triangle$	&	0.276$\blacktriangle$	& 0.411$\blacktriangle$ 	& 0.430$\blacktriangle$				 \\
\hline
MQL$_{dir}$		&		0.247	&	0.308	&	0.192	&	0.303	& 0.420 	& 0.427				  \\
SPUD$_{dir}$	&		0.252$\blacktriangle$	&	0.319$\triangle$	&	0.200$\triangle$	&	0.314$\blacktriangle \bullet$	& 0.431$\blacktriangle \bullet$ 	& 0.445$\blacktriangle \circ$			  \\
\hline
\end{tabular}}  
\end{table}

\begin{table}[!ht] 
%\scriptsize

\tbl{NDCG@20 of SPUD models vs MQL models ($\blacktriangle$ means two-sided t-test $p < 0.01$,  
$\triangle$ means $p < 0.05$) and SPUD models vs DCM-L-T ($\bullet$ means 
two-sided t-test $p  < 0.01$ compared to DCM-L-T, $\circ$ means $p    <   0.05$ 
compared to DCM-L-T).  \label{tab:results_title_ndcg}}{
\centering
\renewcommand{\arraystretch}{1.1}
\setlength\tabcolsep{3pt}
\begin{tabular}{| l || l | l | l | l | l| l | l | l | l |}

\hline	
				&		 \multicolumn{6}{c|}{short queries} 	\\
\hline
				&  	  {robust-04} 		& {trec-8} & {trec-9/10} & 	{gov2} &	{mq-07}	&	{mq-08}	\\

\hline
DCM-L-T			&		0.423	&	0.449	&	0.298	&	0.455	& 0.465 	& 0.495			 \\
%\hline
\hline
MQL$_{jm}$		&		0.385	&	0.356	&	0.220	&	0.379	& 0.458 	& 0.495			 \\
SPUD$_{jm}$		&		0.398 $\blacktriangle$	&	0.384$\triangle$	&	0.243$\triangle$	&	0.418$\blacktriangle$	& 0.474$\triangle$ 	& 0.503		 \\
\hline
MQL$_{dir}$		&		0.423	&	0.466	&	0.309	&	0.470	& 0.488 	& 0.500			  \\
SPUD$_{dir}$		&		0.432 $\blacktriangle $	&	0.477 	&	0.322 	&	0.492$\blacktriangle \bullet$	& 0.500$\blacktriangle \circ$ 	& 0.513$\blacktriangle \circ$			  \\
\hline
\end{tabular}}
\end{table}

Tables \ref{tab:results_desc_map} and \ref{tab:results_desc_ndcg} show
the retrieval  effectiveness (MAP and NDCG@20)  of MQL$_{jm}$ compared
to SPUD$_{jm}$,  and MQL$_{dir}$  compared to SPUD$_{dir}$  for medium
length queries (6-10 terms on average).  Again we can see that on most
of the  test collections  the SPUD models  demonstrate an  increase in
effectiveness for  both MAP  and NDCG@20. All  of these  increases are
significant in the case of SPUD$_{dir}$. For long queries (10-30 terms
on average) we see a similar  trend. A point worth emphasising is that
the  increases in effectiveness  are also  present at  the top  of the
ranked lists as demonstrated by NDCG@20.

The SPUD$_{dir}$ approach outperforms the previous DCM relevance-based
model (DCM-L-T) on most test collections. We have found  that the DCM-L-T  
performs similarly  to MQL$_{dir}$  for short queries on some of the  
smaller collections, but we find that the DCM-L-T approach performs 
quite poorly on the larger {gov2}, {mq-07},
and   {mq-08}  test   collections  and   for  all   medium   and  long
queries. Statistical  significance tests (using a  two-sided t-test 
indicated  by  $\circ$ and $\bullet$)  show  that  the  best
performing SPUD model (SPUD$_{dir}$) outperforms the DCM-L-T approach 
on some collections for short queries and consistently outperforms a tuned
DCM-L-T approach for longer queries. We discuss some possible reasons for 
these results in Section~7.1.

\begin{table}[!ht] 
%\scriptsize

\tbl{MAP of SPUD models vs MQL models  ($\blacktriangle$ means two-sided t-test $p < 0.01$,  
$\triangle$ means $p < 0.05$) and SPUD models vs DCM-L-T ($\bullet$ means 
two-sided t-test $p  < 0.01$ compared to DCM-L-T, $\circ$ means $p    <   0.05$ 
compared to DCM-L-T).\label{tab:results_desc_map}}{
\centering
\renewcommand{\arraystretch}{1.3}
\setlength\tabcolsep{3pt}
\begin{tabular}{| l || l | l | l | l | l| l | l | l | l |}

\hline	
				&		 \multicolumn{5}{c|}{medium length queries} 	\\
\hline
				&  	  {robust-04} 		& {trec-8} & {trec-9/10} & 	{gov2} &	{ohsu}		\\

\hline
DCM-L-T			&		0.266	&	0.296	&	0.181	&	0.256	& 0.255 				 \\
%\hline
\hline
MQL$_{jm}$		&		0.277	&	0.283	&	0.191	&	0.276	& 0.239 				 \\
SPUD$_{jm}$		&		0.280	&	0.291	&	0.203$\blacktriangle \circ$	&	0.299$\blacktriangle \bullet$	& 0.248$\triangle$ 				 \\
\hline
MQL$_{dir}$		&		0.281	&	0.325	&	0.238	&	0.315	& 0.253 				  \\
SPUD$_{dir}$	&		0.289$\blacktriangle \bullet$	&	0.347$\blacktriangle \bullet$	&	0.247$\blacktriangle \bullet$	&	0.329$\blacktriangle \bullet$	& 0.270$\blacktriangle \bullet$				  \\
\hline
\end{tabular}}
\end{table}

\begin{table}[!ht] 

\tbl{NDCG@20 of SPUD models vs MQL models  ($\blacktriangle$ means two-sided t-test $p < 0.01$,  
$\triangle$ means $p < 0.05$) and SPUD models vs DCM-L-T ($\bullet$ means 
two-sided t-test $p  < 0.01$ compared to DCM-L-T, $\circ$ means $p    <   0.05$ 
compared to DCM-L-T).\label{tab:results_desc_ndcg}}{
\centering
\renewcommand{\arraystretch}{1.1}
\setlength\tabcolsep{3pt}
\begin{tabular}{| l || l | l | l | l | l| l | l | l | l |}

\hline	
				&		 \multicolumn{5}{c|}{medium length queries} 	\\
\hline
				&  	  {robust-04} 		& {trec-8} & {trec-9/10} & 	{gov2} &	{ohsu}		\\

\hline
DCM-L-T			&		0.435	&	0.436	&	0.318	&	0.401	& 0.396 				 \\
%\hline
\hline
MQL$_{jm}$		&		0.455	&	0.412	&	0.329	&	0.431	& 0.397 				 \\
SPUD$_{jm}$		&		0.456	&	0.440$\triangle$	&	0.344$\triangle \bullet$	&	0.463$\blacktriangle \bullet$	& 0.391 				 \\
\hline
MQL$_{dir}$		&		0.465	&	0.478	&	0.393	&	0.484	& 0.399				  		\\
SPUD$_{dir}$	&		0.479$\blacktriangle \bullet$	&	0.500$\blacktriangle \bullet$	&	0.403$\triangle \bullet$ &	0.502$\blacktriangle \bullet$	& 0.415$\blacktriangle \circ$				  \\
\hline
\end{tabular}}
\end{table}

\begin{table}[!ht] 
%\scriptsize

\tbl{MAP of SPUD models vs MQL models  ($\blacktriangle$ means two-sided t-test $p < 0.01$,  
$\triangle$ means $p < 0.05$) and SPUD models vs DCM-L-T ($\bullet$ means 
two-sided t-test $p  < 0.01$ compared to DCM-L-T, $\circ$ means $p    <   0.05$ 
compared to DCM-L-T).\label{tab:results_narr_map}}{
\centering
\renewcommand{\arraystretch}{1.3}
\setlength\tabcolsep{3pt}
\begin{tabular}{| l || l | l | l | l | l| l | l | l | l |}

\hline	
				&		 \multicolumn{4}{c|}{long queries} 	\\
\hline
				&  	  {robust-04} 		& {trec-8} & {trec-9/10} & 	{gov2} 		\\

\hline
DCM-L-T			&		0.239	&	0.225	&	0.181	&	0.235					 \\
%\hline
\hline
MQL$_{jm}$		&		0.284	&	0.269	&	0.211	&	0.265					 \\
SPUD$_{jm}$		&		0.288$\triangle \bullet$	&	0.269 $\bullet$	&	0.206 $\circ$	&	0.285 $\blacktriangle \bullet$	 				 \\
\hline
MQL$_{dir}$		&		0.283	&	0.283	&	0.248	&	0.296
	 				  \\
SPUD$_{dir}$	&		0.296$\blacktriangle \bullet$	&	0.314$\blacktriangle \bullet $ 	&	0.254 $\bullet $	&	0.323 $\blacktriangle \bullet$	 				  \\
\hline
\end{tabular}}
\end{table}

\begin{table}[!ht] 

\tbl{NDCG@20 of SPUD models vs MQL models  ($\blacktriangle$ means two-sided t-test $p < 0.01$,  
$\triangle$ means $p < 0.05$) and SPUD models vs DCM-L-T ($\bullet$ means 
two-sided t-test $p  < 0.01$ compared to DCM-L-T, $\circ$ means $p    <   0.05$ 
compared to DCM-L-T).\label{tab:results_narr_ndcg}}{
\centering
\renewcommand{\arraystretch}{1.1}
\setlength\tabcolsep{3pt}
\begin{tabular}{| l || l | l | l | l | l| l | l | l | l |}

\hline	
				&		 \multicolumn{4}{c|}{long queries} 	\\
\hline
				&  	  {robust-04} 		& {trec-8} & {trec-9/10} & 	{gov2} 	\\

\hline
DCM-L-T			&		0.404	&	0.346	&	0.318	&	0.400					 \\
%\hline
\hline
MQL$_{jm}$		&		0.469	&	0.430	&	0.354	&	0.535					 \\
SPUD$_{jm}$		&		0.476 $\bullet$	&	0.431 $\bullet$	&	0.356 $\bullet$	&	0.541 $\bullet$	 				 \\
\hline
MQL$_{dir}$		&		0.467	&	0.446	&	0.406	&	0.572	 				  \\
SPUD$_{dir}$	&		0.483$\blacktriangle \bullet$	&	0.475$\triangle \bullet$	&	0.409 $\bullet$	&	0.599 $\blacktriangle \bullet$  \\
\hline
\end{tabular}}
\end{table}

\begin{figure}[!ht] 
\begin{center}
$\begin{array}{c c c}
	\includegraphics[height=3.5cm,width=4cm]{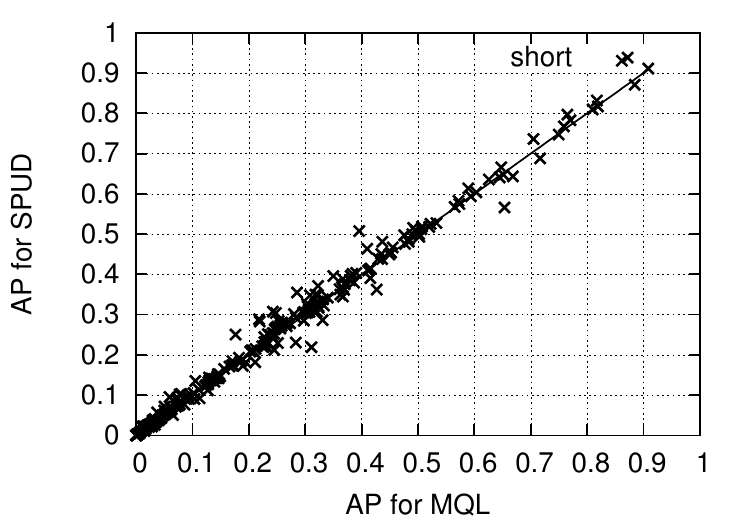} &
	\includegraphics[height=3.5cm,width=4cm]{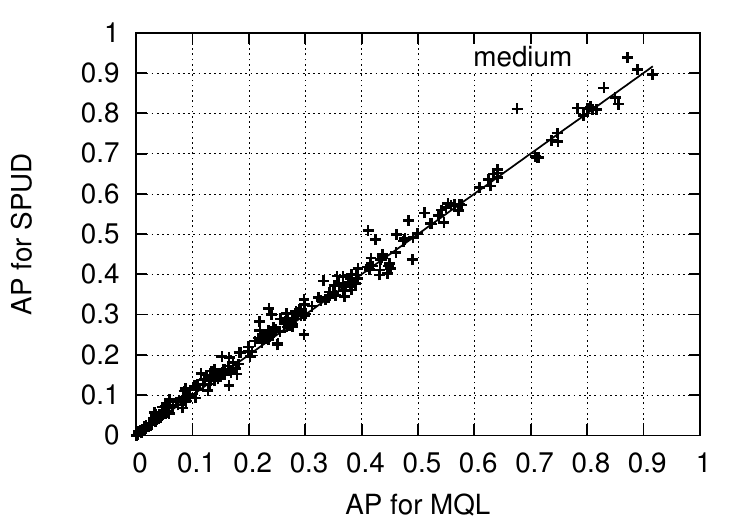} &
	\includegraphics[height=3.5cm,width=4cm]{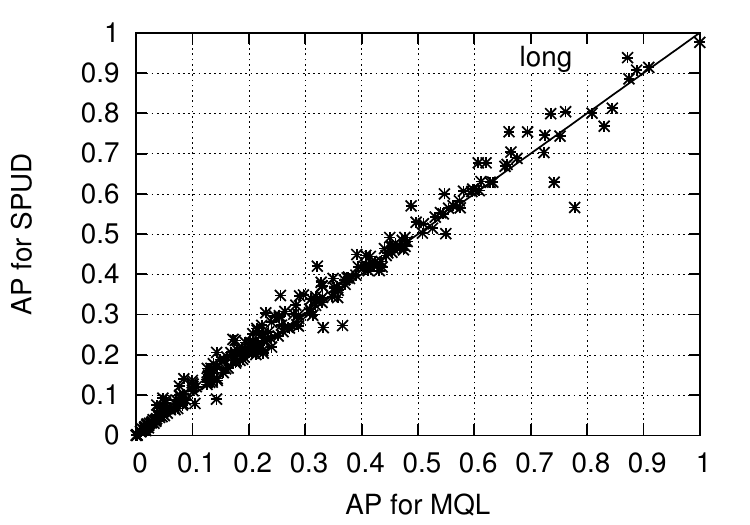} \\
	
	\includegraphics[height=3.5cm,width=4cm]{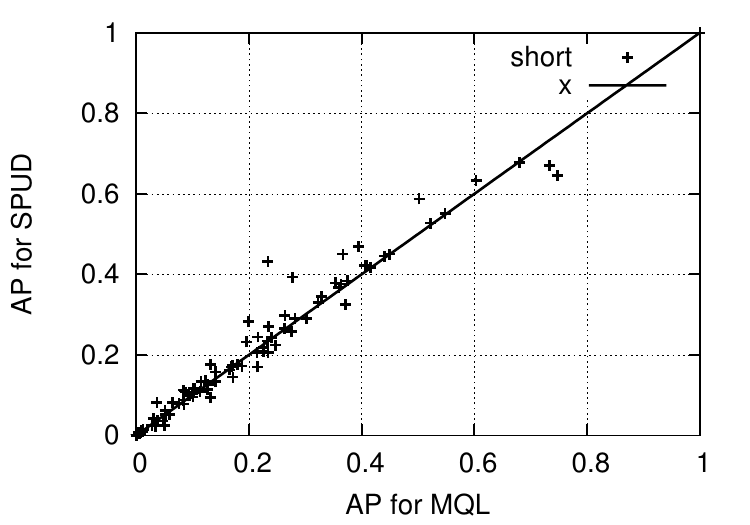} &
	\includegraphics[height=3.5cm,width=4cm]{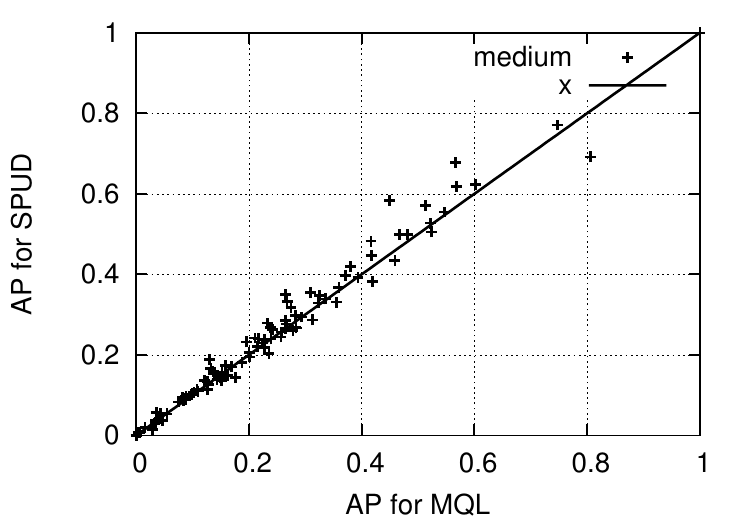} &
	\includegraphics[height=3.5cm,width=4cm]{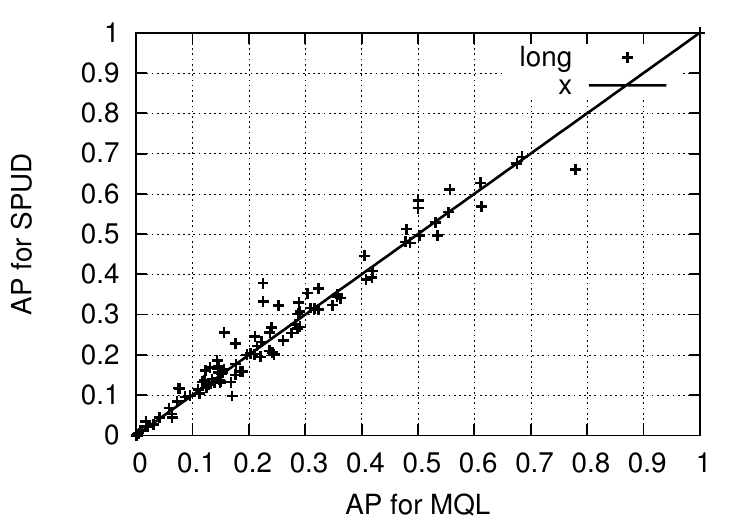} \\	
\end{array}$
\caption{Average precision of all  short, medium, and long queries for
  MQL$_{dir}$  vs  SPUD$_{dir}$   on  {robust-04}  dataset  (top)  and
  trec-9/10 (bottom)}
\label{fig:query_analysis_1}
\end{center}
\end{figure}

Fig.\ref{fig:query_analysis_1} shows the performance of MQL$_{dir}$ vs
SPUD$_{dir}$  for each query  on two  separate test  collections. This
query specific  analysis indicates that SPUD$_{dir}$ is  robust on all
ranges of queries (from easy  to difficult). For longer queries on the
robust-04 dataset, there are one  or two high performing queries which
drop over 0.1 in average precision. However, in general there are very
few queries  which severely under perform compared  to MQL$_{dir}$. On
the  trec-9/10 web documents,  the increase  in performance  is stable
across all types of queries for all query lengths.

\subsection{Robustness}

The second experiment evaluates the robustness of the SPUD models with
respect to different parameter  settings. In addition, we evaluate the
retrieval effectiveness  of the SPUD$_{dir}$ model  when the parameter
$\mu'$ is  derived from the  estimated parameter $m_c$  using Newton's
method \cite{elkan06}.

\subsubsection{Robustness Results}

\begin{figure}[!ht] 
\begin{center}
$\begin{array}{c c }
	\includegraphics[height=4cm,width=4.5cm]{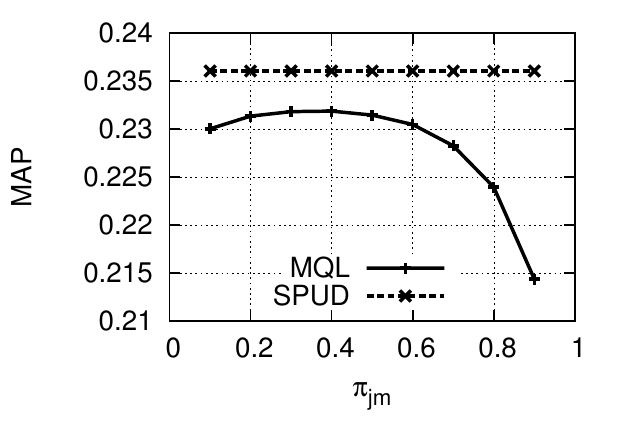} &
	\includegraphics[height=4cm,width=4.5cm]{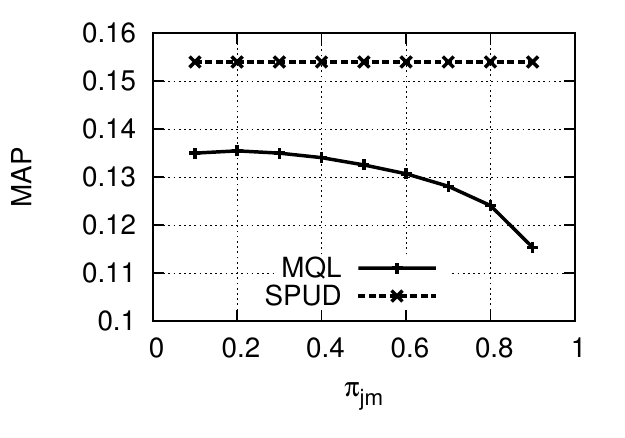} 
\end{array}$
\caption{Robustness  comparison  of   MQL$_{jm}$  and  SPUD$_{jm}$  on
  robust-04 (left) and {trec-9/10} (right) for short queries}
\label{fig:rbow_jm_tuning}
\end{center}
\end{figure}

Fig.~\ref{fig:rbow_jm_tuning} shows the performance of MQL$_{jm}$ over
different   tuning  parameter   values  (i.e.   $\pi_{jm}$)   and  the
performance  of the SPUD$_{jm}$  model. We  can see  that SPUD$_{jm}$,
which  has  no  free  parameters,  outperforms  SPUD$_{jm}$  over  all
parameter  values. This trend  is consistent  on all  test collections
used here.

\begin{figure}[!ht] 
\begin{center}
$\begin{array}{c                                                      }
    \includegraphics[height=4.5cm,width=5.5cm]{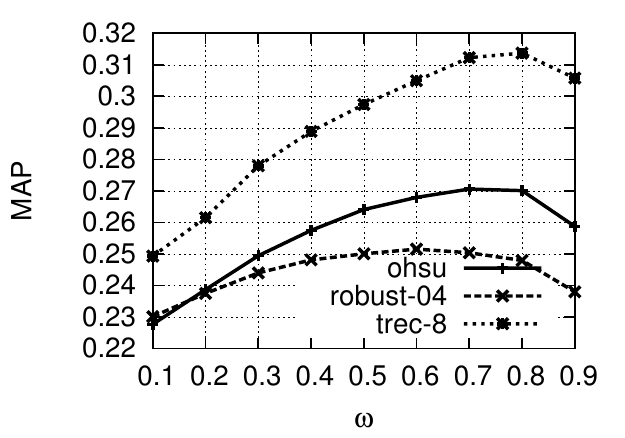}
\end{array}$
\caption{Tuning  of  $\omega$ in  SPUD$_{dir}$  on robust-04  (short),
  trec-8  (short),  and  ohsu   (medium)  collections  when  $m_c$  is
  estimated using Eq.~(\ref{eq:estimate_w})}
\label{fig:gamma_tuning}
\end{center}
\end{figure}

For the  SPUD$_{dir}$ function, we  can estimate $m_c$  using Newton's
method  as   outlined  in  Eq.~(\ref{eq:estimate_w})   given  the  $n$
documents as data.  We found that  an initial value of $m_c = 200$ was
suitable  so that  the process  converged within  20  iterations. This
computation can be done off-line  and we used the resulting setting of
$m_c$ to  estimate $\mu'$ by  tuning the hyperparameter $\omega$  to a
fixed  value.   We  set  $\omega  =  0.8$  which  is  demonstrated  in
Fig.~\ref{fig:gamma_tuning}      as     a      reasonable     setting.
Fig.~\ref{fig:gamma_tuning}  shows   the  performance  (MAP)   of  the
SPUD$_{dir}$  model for  different values  of $\omega$  when  $m_c$ is
estimated using Newton's method. The relationship between $\omega$ and
$\mu'$  in Eq.~(\ref{eq:rbow_mixing_param}) essentially  suggests that
$\mu'  =  4 \cdot  m_c$  is a  suitable  parameter  value for  $\mu'$.
Although $m_c$ is the only  parameter that is expensive to estimate in
the SPUD$_{dir}$ model,  it is practically feasible to  do so offline.
When the  parameter $\mu'$  is computed  in this way  (i.e. $\mu'  = 4
\cdot m_c$),  we denote  this SPUD$_{est_{\mu'}}$ in  the experimental
results and figures that follow.

%For the SPUD$_{dir}$ model, the tuning parameter $\mu'$ can be estimated 
%by first estimating $m_c$ and then by tuning $\omega$. The $m_c$ value in 
%the SPUD$_{dir}$ model can be estimated using all $n$ documents in the collection 
%and Newton's method. The estimation process for $m_c$ converges within 20 iterations.
%Moreover, to completely specify the SPUD$_{dir}$ model we need to find an appropriate
%mixing parameter $\omega$. The $\omega$ parameter specifies the weight of the background 
%DCM compared to the document DCM. 

\begin{figure}[!ht] 
\begin{center}
$\begin{array}{c c }
	\includegraphics[height=4cm,width=4.5cm]{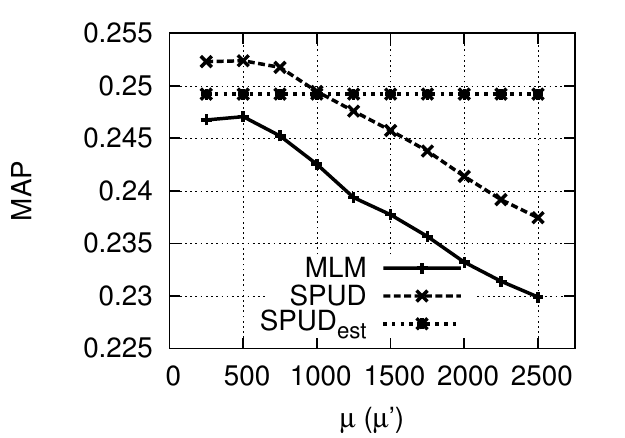} &
	\includegraphics[height=4cm,width=4.5cm]{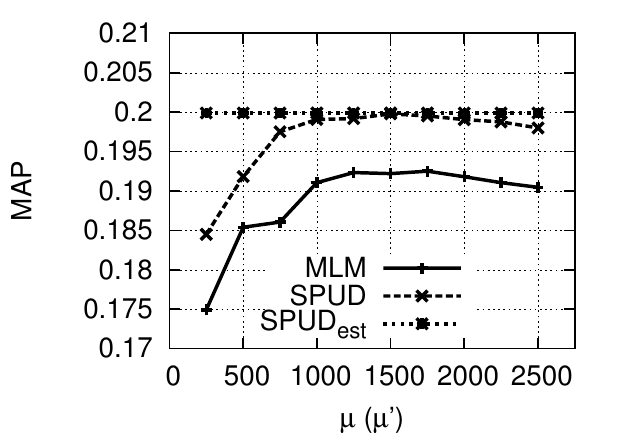}
\end{array}$
\caption{Robustness  of SPUD$_{dir}$  and  MQL$_{dir}$ over  different
  values  of  the tuning  parameter  $\mu$  (or  $\mu'$) on  robust-04
  (short) and {trec-9/10} (short) respectively}
\label{fig:rbow_dir_tuning}
\end{center}
\end{figure}

Fig.~\ref{fig:rbow_dir_tuning}  shows  the performance of  MQL$_{dir}$,
SPUD$_{dir}$,   and  SPUD$_{est_{\mu'}}$   over   different  parameter
settings on a number of test collections. We can see that SPUD$_{dir}$
outperforms MQL$_{dir}$ over all parameter values. We can see that the
parameter $\mu'$ is  as robust as the parameter $\mu$,  as it tends to
follow the same trend.

More importantly we see that near optimal effectiveness can usually be
achieved  by using the  automatically estimated  value of  $m_c$ found
using Newton's method. This is rather encouraging as it means that the
setting of  $\omega = 0.8$ is  robust and that we  can effectively and
safely eliminate from SPUD$_{dir}$ the free parameters. In particular,
this  automatic optimal  estimation can  be  seen when  we examine  in
Fig.~\ref{fig:rbow_dir_tuning}   the  {trec-9/10}   collection  (which
contains long Web documents) and the {robust-04} collection (which has
shorter  documents).  For the  {robust-04}  collection, the  retrieval
effectiveness  decreases  sharply  when  $\mu'$ becomes  greater  than
1000. On the other hand, for the {trec-9/10} the effectiveness is more
stable when $\mu'$ is greater  than 1000. One probable reason for this
is that  the average length of  the documents in  those collections is
very     different.    However,     the     automatically    estimated
SPUD$_{est_{\mu'}}$ is close to optimal on both collections.

Table~\ref{tab:rbow_dir_optimal}            reinforces            this
observation.      Table~\ref{tab:rbow_dir_optimal}      shows      the
characteristics of the average  length of documents in the collections
and the  value of $m_c$ that  is estimated on each  collection. We can
also see that $m_c$ is correlated with the lengths of documents in the
collections. Furthermore, in  the same table we can  see that close to
the  optimal effectiveness  is  possible by  setting $\omega=0.8$  for
SPUD$_{est_{\mu'}}$. This  is because $m_c$  is essentially performing
the tuning on  a per collection basis. The parameter  $m_c$ has a very
intuitive interpretation as the initial mass of the background P\'olya
urn.

\begin{table}[!ht] 

\tbl{MAP comparison  of SPUD$_{dir}$  model for well-tuned  $\mu'$ and
  SPUD$_{est_{\mu'}}$ which  uses an automatically  estimated value of
  $\mu'$           \label{tab:rbow_dir_optimal}}{           \centering
  \renewcommand{\arraystretch}{1.1} \setlength\tabcolsep{3pt}
\begin{tabular}{| l || l | l | l | l | l | l | l | l |}

\hline
						&  {robust-04}	& {trec-8} & {trec-9/10} & 	{ohsu}	&	gov2 &	mq-07 &	mq-08 \\

\hline
\hline
 
$|\vec{d}|_{avg}$		& 162	& 242 & 157 & 68 & 181 & 181 & 181 \\
\hline
$|d|_{avg}$				& 265	& 558 & 344 & 104  & 529 & 529 & 529 \\
\hline
$\hat{m_c}$				& 258	&  421 & 326 & 112 & 234 & 234 & 234 \\
\hline
$\hat{u'} = 4 \cdot \hat{m_c}$	& 1034	& 1688 & 1308 & 448 & 936 & 936 & 936 \\
\hline
	&	 \multicolumn{7}{c|}{short queries} \\
\hline
SPUD$_{dir}$				& 0.252	& 0.319 & 0.200 & n/a & 0.314 & 0.431 & 0.445 \\
SPUD$_{est_{\mu'}}$		& 0.249	& 0.320 & 0.199 & n/a & 0.314 & 0.429 & 0.443 \\
\hline
	&	\multicolumn{7}{c|}{medium queries} \\
\hline
SPUD$_{dir}$				& 0.289	& 0.347 & 0.247 & 0.270 & 0.332 & n/a & n/a \\
SPUD$_{est_{\mu'}}$		& 0.287	& 0.344 & 0.246 & 0.270 & 0.329 & n/a & n/a \\

\hline
	&	\multicolumn{7}{c|}{long queries} \\
\hline
SPUD$_{dir}$				& 0.296	& 0.314 & 0.254 & n/a & 0.323 & n/a & n/a \\
SPUD$_{est_{\mu'}}$		& 0.295	& 0.307 & 0.255 & n/a & 0.322 & n/a & n/a \\

\hline
\end{tabular}}
\end{table}

\subsection{Analysis of Retrieval Model Aspects}

In   this  third  experiment   we  aim   to  evaluate   the  retrieval
effectiveness of  the new background model  (i.e. $\bm{\alpha}_c$) and
the new smoothing methods in the SPUD model separately in a piece-wise
fashion. We gradually adapt  parts of the multinomial query likelihood
functions  until  the  SPUD  retrieval functions  are  comprised.  The
experiment pinpoints  the parts of  the SPUD retrieval  functions that
lead to changes in retrieval effectiveness. This piece-wise adaptation
provides evidence that the individual retrieval intuitions outlined in
Section~5.2  are valid.  Furthermore, we  conduct an  analysis  of the
retrieval characteristics of the best performing methods.

\subsubsection{Results of the Analysis of Retrieval Model Aspects}

Table~\ref{tab:decomposition},  which  also contains  a  column for  a
\emph{hybrid} model,  outlines the parameter values  for the functions
used  in this  experiment. Essentially,  this  \emph{hybrid} retrieval
function differs  from the SPUD  retrieval functions only in  the fact
that  it uses  different  parameter estimates  for $\lambda_{jm}$  and
$m_d$  that  effect the  smoothing  for  SPUD$_{jm}$ and  SPUD$_{dir}$
respectively.  The  changes to  these  parameter  estimates makes  the
\emph{hybrid} model closer to the multinomial retrieval functions. The
only difference  between the MQL  and $hybrid$ model is  that $hybrid$
uses    the    expected   multinomial    of    the   background    DCM
(i.e.  $\bm{\theta}_c'$ in  Eq.~\ref{eq:back_prob}) as  its background
model.

% somewhere in your document's body:
\begin{table}[!ht] 
\tbl{Decomposition of Retrieval Functions \label{tab:decomposition}}{
\centering
\renewcommand{\arraystretch}{1.3}
\setlength\tabcolsep{10pt}
\begin{tabular}{|l|l|l|l|l|}

\hline
		 			&  		 & MQL 		& $hybrid$  & SPUD	\\
Smoothing 			& Colour & Multinomial 	& DCM 		& DCM	\\
\hline
\hline
Jelinek-Mercer (jm) 	& Blue 	& $\pi_{jm} = 0.2$  	& $\lambda_{jm} = 0.2$			&	$\lambda_{jm} = |\vec{d}|/|d|$	\\
\hline
Dirichlet (dir) 		& Red & $\mu = 2000$ 		& $\mu' = 2000$, $m_d = |d|$  	& 	$\mu' = 2000$, $m_d = |\vec{d}|$	\\
\hline

\end{tabular}}
\end{table}

\begin{figure}[!ht] 
\begin{center}
$\begin{array}{c c c}
   	\includegraphics[height=4cm,width=5cm]{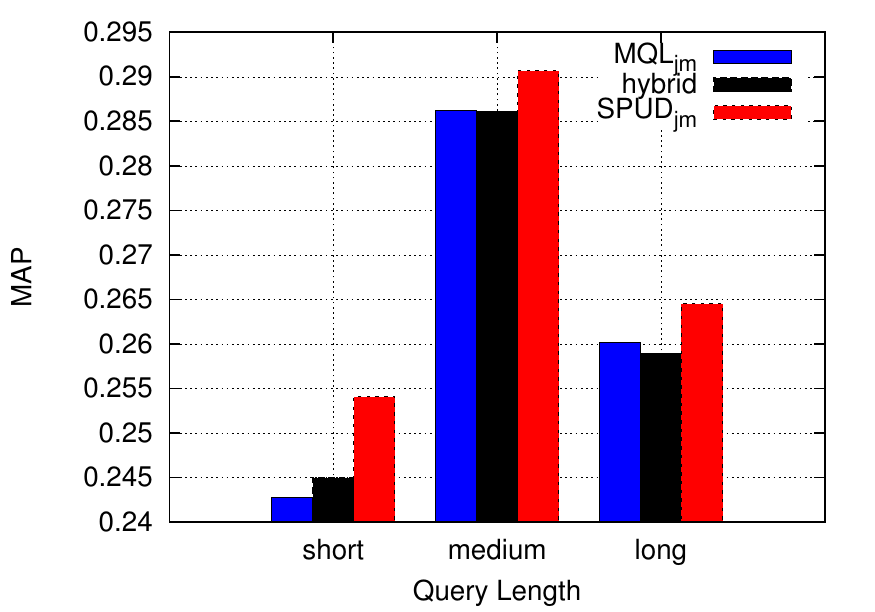} &    
	\includegraphics[height=4cm,width=5cm]{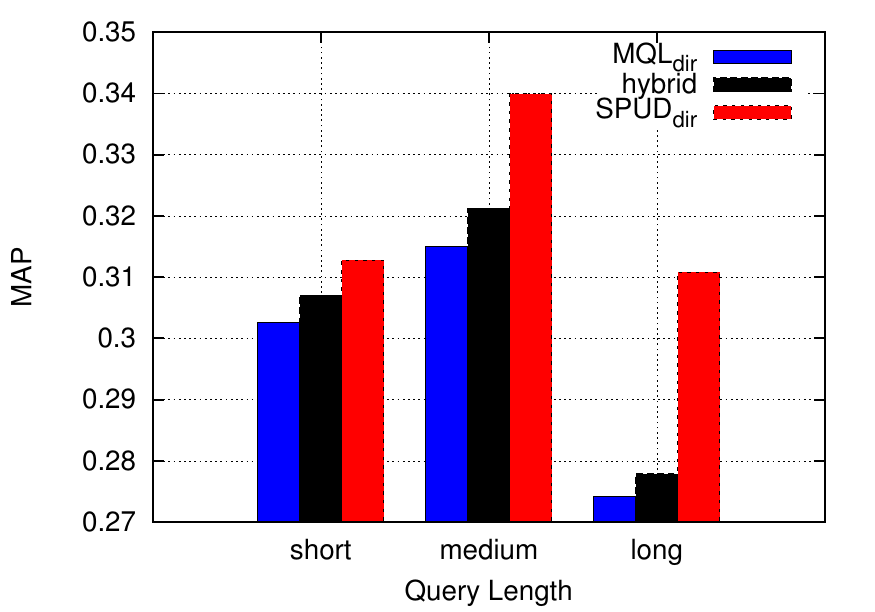} \\
	\includegraphics[height=4cm,width=5cm]{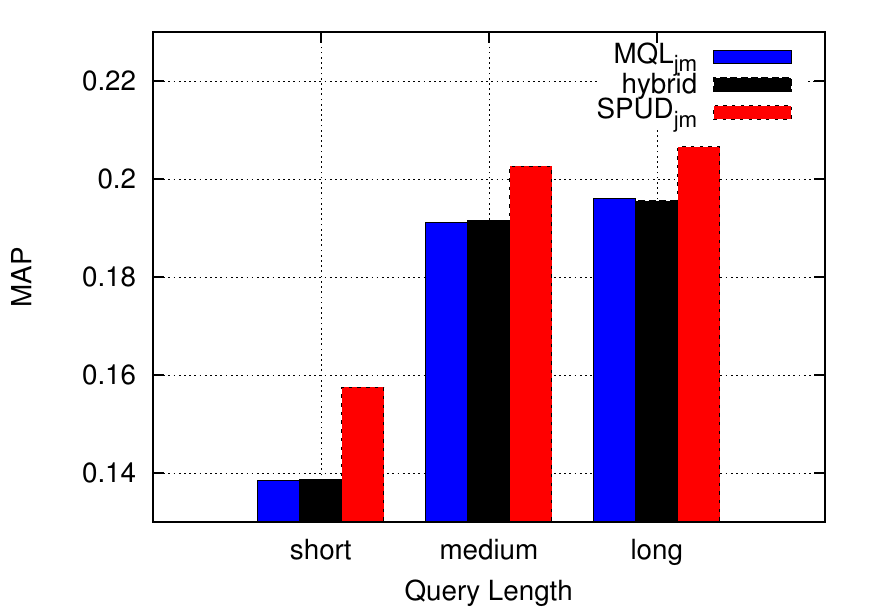} &
	\includegraphics[height=4cm,width=5cm]{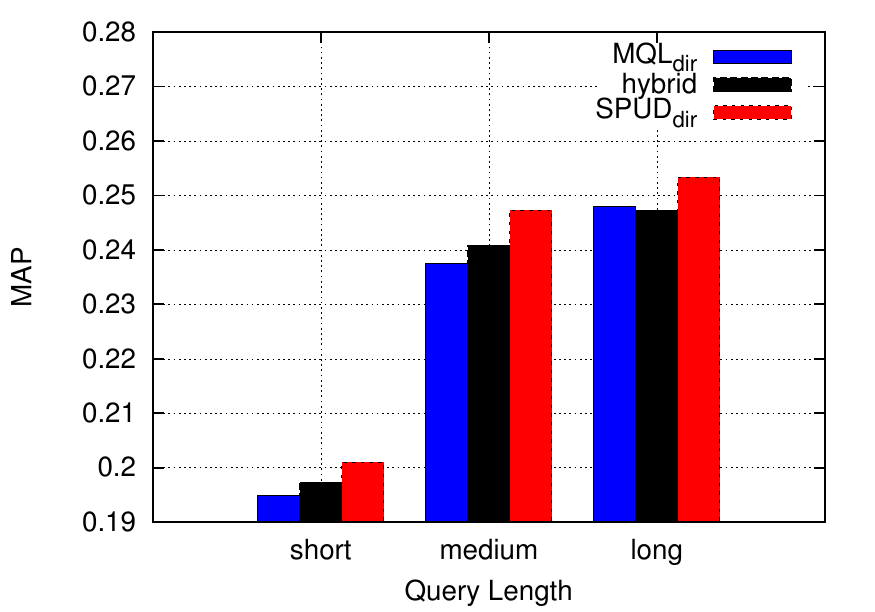} 
\end{array}$
\caption{Analysis  of performance  gains from  different parts  of the
  SPUD retrieval  models for trec-8  (top) and trec-9/10  (bottom) for
  short,  medium, and  long  queries. Models  that use  Jelinek-Mercer
  smoothing are on the left-hand,  side while those that use Dirichlet
  smoothing are on the right-hand side. }
\label{fig:rbow_dir_gains}
\end{center}
\end{figure}

Fig.~\ref{fig:rbow_dir_gains} shows the effectiveness of the functions
that use Jelinek-Mercer smoothing  (left-hand side) and those that use
a  type   of  Dirichlet  smoothing  (right-hand  side)   on  two  test
collections. In general, the use  of the new background DCM model aids
retrieval as we can see  an increase in effectiveness for the $hybrid$
(in black)  retrieval functions over  the MQL functions (in  blue). We
note that the  magnitude of the difference is small,  and that in some
cases the performance decreases slightly. In general, the introduction
of a document boundary into  the estimation of the background language
model is more effective for SPUD$_{dir}$ than for SPUD$_{jm}$.

However,  the different  smoothing techniques  introduced in  the SPUD
model yield a greater increase  in performance i.e. when comparing the
SPUD  function (in  red)  to  the $hybrid$  function  (in black).  The
smoothing in  the SPUD model, amongst other  factors, affects document
length    normalisation   and    improves    retrieval   effectiveness
substantially.    The    results   in    Fig.~\ref{fig:rbow_dir_gains}
demonstrate\footnote{Results  on other test  collections used  in this
  work     are      consistent     with     those      reported     in
  Fig.~\ref{fig:rbow_dir_gains}.}     that    the     new    retrieval
characteristics  brought about  by  both the  background  DCM and  the
document DCM positively influence retrieval effectiveness. The results
of these  experiments further validate  the use of  the DCM as  a more
plausible  document model than  the multinomial.  This is  because the
changes  to  the  query-likelihood   retrieval  method  that  the  new
background and new document model  bring about, increase the
retrieval effectiveness for SPUD$_{dir}$ method over MQL.

\begin{figure}[!ht] 
\begin{center}
$\begin{array}{c c c}
   	\includegraphics[height=4cm,width=5cm]{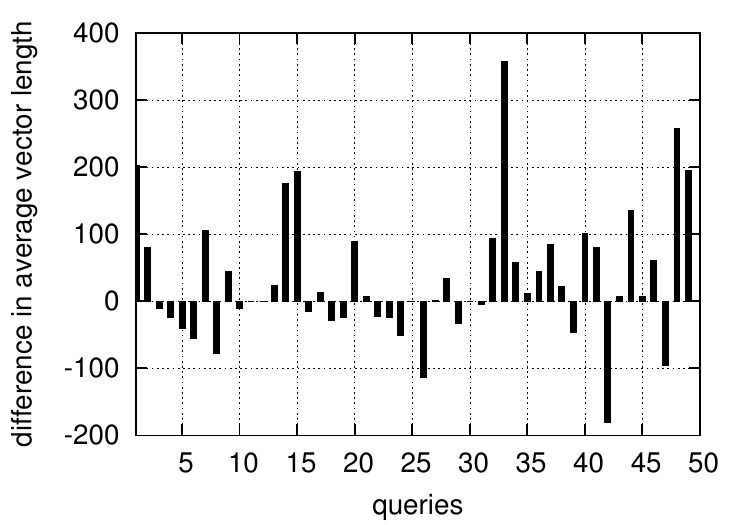} &    
	\includegraphics[height=4cm,width=5cm]{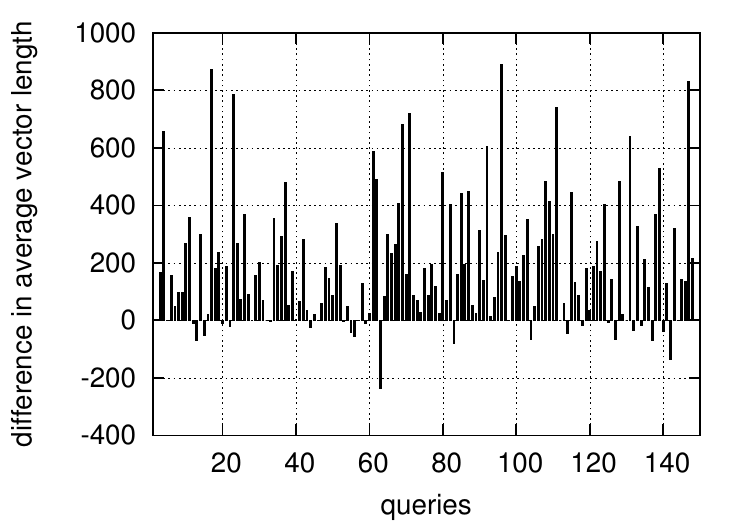} \\
\end{array}$
\caption{Difference in  average vector length  of the top  20 returned
  documents for SPUD$_{dir}$ and MQL$_{dir}$ on trec-8 (left) and gov2
  (right) web collections for short queries.}
\label{fig:vector_lengths}
\end{center}
\end{figure}

Previously  in Section {5.2.2}  we analysed  the lengths  of documents
retrieved  in   the  top  1000  documents  by   both  MQL$_{dir}$  and
SPUD$_{dir}$. We  found that SPUD$_{dir}$ was more  likely to retrieve
longer documents. We now look at the length characteristics of the top
20 documents  returned per query by both  SPUD$_{dir}$ and MQL$_{dir}$
to  determine  if  the  differences  in  length  are  correlated  with
increased    performance     in    terms    of     NDCG@20.    Firstly
Fig.\ref{fig:vector_lengths}  confirms  that  on average  SPUD$_{dir}$
retrieves documents  with a longer  vector length than  MQL$_{dir}$ in
the top 20.  Table~\ref{tab:correlation} shows the correlation between
the differences in  average length and the differences  in NDCG in the
top   20   documents   across   a  number   of   representative   test
collections. We  report a small but  insignificant correlation between
the  increase in  average vector  length and  query  effectiveness (as
measured by  NDCG@20). Although this correlation  analysis is somewhat
inconclusive, we  can confirm  that on average  SPUD$_{dir}$ retrieves
documents with a longer vector length (i.e. greater number of distinct
terms), and  that the overall evidence  seems to suggest  that this is
leading to increase effectiveness.

\begin{table}[!ht] 
\tbl{Linear correlation of $\Delta$  average document length in top 20
  and   $\Delta$   NDCG@20   over   short   queries   sets   for   Web
  collections\label{tab:correlation}}{                       \centering
  \renewcommand{\arraystretch}{1.3} \setlength\tabcolsep{10pt}
\begin{tabular}{|c||l|l|l|}

\hline
avg\_len 		&  {trec-8} & {trec-9/10} & 	gov2\\
\hline
$|d|$			&	0.0525		&	0.0462		&	-0.0161		\\
$\vec{|d|}$		&	0.0838		&	0.0622		&	0.0216		\\

\hline

\hline
\end{tabular}}
\end{table}

\subsection{Pseudo-Relevance Feedback}

Finally,  we evaluate the  SPUD model  in a  pseudo-relevance feedback
setting. Pseudo-relevance feedback is  a useful approach for expanding
short  queries when  the  user  has not  entered  a sufficiently  long
query. In  essence, the pseudo-relevance model is  responsible for the
selection and  weighting of  candidate expansion query-terms  from the
top  $k$  documents  of  an   initial  retrieval  run.  We  adapt  the
state-of-the-art  RM3  \cite{umass4,diaz06}  approach  to  select  and
weight terms  according to  the SPUD$_{dir}$ retrieval  approach.  The
pseudo-relevance  model based  on  SPUD$_{dir}$ is  estimated from  an
initial ranking as follows:

\begin{equation} \label{eq:rbow_rm3}
p(t|q_{e}) = \sum_{\bm{\alpha}_{dm} \in \hat{R_{\alpha}}}  p(t|\bm{\alpha}_{dm})  \frac{p(q|\mathcal{M}_d = \mathop{\mathbb{E}} [\bm{\theta}_{dm} | \bm{\alpha}_{dm}])}{ \sum_{\bm{\alpha'}_{dm} \in \hat{R_{\alpha}}} p(q|\mathcal{M}_d = \mathop{\mathbb{E}} [\bm{\theta}_{dm} | \bm{\alpha'}_{dm}]) }
\end{equation}
where $\hat{R_{\alpha}}$ is the set of pseudo-relevant document models
(i.e.  it  is the  top $k$ document  models from an  initial retrieval
run).      If      we     replace     $p(t|\bm{\alpha}_{dm})$     with
$p(t|\bm{\theta}_{dm})$         and        $p(q|\mathcal{M}_d        =
\mathop{\mathbb{E}}[\bm{\theta}_{dm} | \bm{\alpha}_{dm}])$ with $p(q |
\bm{\theta}_{dm})$  in Eq.~(\ref{eq:rbow_rm3}),  we  recover RM3.  The
final  query  model  is  then  estimated by  linearly  smoothing  this
estimated  relevance model  $p(t|q_{e})$  with the  original query  as
follows:

\begin{equation} \label{eq:rm3_smoothing}
p(t|q') =  \tau \cdot p(t|q) + (1 - \tau) \cdot p(t|q_{e})
\end{equation}
where $\tau$ controls  the weight of the initial  query. The new query
model is  then used  to query the  corpus using the  initial retrieval
method  (i.e.  SPUD$_{dir}$). We  set  the  number of  pseudo-relevant
documents $k=20$ and generate a pseudo-relevance model of 50 terms. We
smooth  the pseudo-relevance model  with the  original query  model by
setting $\tau=0.5$. The parameter $u'$ (and $u$ in MQL$_{dir}$) is set
to  2000 during  ranking and  is set  to 0  only during  the expansion
step.  These expansion parameters  settings are  set according  to the
literature    \cite{umass4,lv09:3,lv10}.     We    note    that    the
pseudo-relevance  model here  does not  follow a  DCM  relevance model
(i.e. we do not treat all relevant documents as being drawn from a DCM
relevance model), but  is simply an adaptation of  the RM3 model which
we  refer to  as {\bf  PURM}\footnote{Essentially, the  PURM expansion
  model with $u'=0$ only differs from  RM3 with $u=0$ in the fact that
  the document  retrieval score used  to weight the expansion  term is
  different.  Therefore, we would  expect only  a small  difference in
  effectiveness.}. We only use  short title queries in this experiment
as  are the types  of queries  to which  query expansion  is typically
applied \cite{carpineto12}.

\subsubsection{Pseudo-Relevance Feedback Results}

Table~\ref{tab:pseudo_results}    shows    the    results    of    the
pseudo-relevance feedback  experiment. Firstly,  we can see  that when
the SPUD$_{dir}$ approach is used as the retrieval method with the RM3
expansion approach, it leads to a significant improvement over the MQL
approach.  This   is  encouraging,  but  hardly   surprising,  as  the
SPUD$_{dir}$ approach has a more effective initial retrieval. However,
when the retrieval  method is static, and only  the expansion approach
is   allowed  to  vary,   the  PURM   approach  outperforms   the  RM3
approach. The  absolute increase in  effectiveness when using  the new
PURM expansion approach is  quite low, but nevertheless is significant
on  trec-8 and  gov2.  This low  increase  in effectiveness  is to  be
expected as the only difference between the RM3 expansion approach and
the PURM approach  (when $u$ and $u'$  are set to 0) is  that the PURM
approach uses the SPUD retrieval  score to weight terms, while the RM3
approach uses  the MQL retrieval score. Overall,  while this validates
that  the  SPUD$_{dir}$ document  retrieval  score  is  useful in  the
expansion  step  of pseudo-relevance  expansion  approaches, the  main
increase   in  effectiveness   comes  from   the  better   ranking  of
SPUD$_{dir}$ compared to MQL$_{dir}$.

A  point  worth  noting  is  that  the  performance  of  the  feedback
approaches on the {mq-07} and  {mq-08} test collections are worse than
for the  initial retrieval  run (no expansion).  It has  been reported
that  pseudo-relevance  feedback  varies  depending on  the  type  and
quality  of the  test collection  with  results showing  little or  no
improvement  when  using  parts   of  the  million  query  track  data
(i.e. {mq-07} and {mq-08}) \cite{meij11}. One possible reason for this
is  that  during  the  creation   of  the  {mq-07}  and  {mq-08}  test
collections  a shallow  pool depth  was used  in order  to  judge more
queries  than  is  usual  for trec  collections.  As  pseudo-relevance
feedback tends to increase average precision by increasing recall, the
lower  number  of  judged   documents  for  the  million  query  track
collections  could affect  the  natural behaviour  of query  expansion
approaches on this collection.

\begin{table}[!ht] 

\tbl{MAP of pseudo-relevance feedback approaches of SPUD$_{dir}$-PURM,
  SPUD$_{dir}$-RM3,   and   MQL$_{dir}$-RM3  ($\blacktriangle$   means
  two-sided  t-test  $p <  0.01$  compared  to MQL$_{dir}$-RM3,  while
  $\triangle$ means $p <  0.05$ compared to MQL$_{dir}$-RM3. $\bullet$
  means  two-sided t-test  $p  < 0.01$  compared to  SPUD$_{dir}$-RM3,
  $\circ$    means   $p    <   0.05$    compared   to    compared   to
  SPUD$_{dir}$-RM3.)      \label{tab:pseudo_results}}{      \centering
  \renewcommand{\arraystretch}{1.1} \setlength\tabcolsep{3pt}
\begin{tabular}{| l | l ||  l | l | l | l | l | l |}

\hline	
	\multicolumn{2}{|c|}{Methods}				&		 \multicolumn{6}{|c|}{short queries} 	\\
\hline
Ranking 	& Expansion 	& 	 {robust-04} 	&	{trec-8} & {trec-9/10} & {gov2} 	&	{mq-07} & {mq-08} \\

\hline
MQL$_{dir}$  & None		 	&	0.232			&	0.308		&  	0.191	&	0.303		&	0.428		&	0.440			\\			
\hline
MQL$_{dir}$  & RM3		 	&	0.258			&	0.322		& 	0.212	&	0.308		&	0.395		&	0.417			\\			
SPUD$_{dir}$ & RM3			&	0.265$\blacktriangle$			&	0.338$\blacktriangle$		& 	0.218$\blacktriangle$	&	0.319		&	0.404		&	0.428			\\		
SPUD$_{dir}$ & PURM			&	0.266$\blacktriangle$			&	0.340$\blacktriangle \bullet$	& 	0.220$\blacktriangle$	&	0.324 $\blacktriangle \circ$	&	0.408$\blacktriangle$ 		&	0.429$\blacktriangle$			\\		
\hline
\end{tabular}}
\end{table}

\section{Discussion}

In this  section we  discuss the main  findings, limitations,  and the
broader impact of this work.

\subsection{Comparison With Previous Work}

The  results of  experiments in  Section~{6.3.1} have  shown  that the
SPUD$_{dir}$  method   significantly  outperforms  the   {DCM-L-T}  of
\citet{xu08}. In particular, the effectiveness of {DCM-L-T} for longer
queries, which was not presented in the original work, is particularly
poor.  The  manner  in  which  the  initial  query  is  used  in  that
relevance-based  model  leads  to  a non-linear  query  term-frequency
aspect.  This is  likely  to affect  the  retrieval effectiveness  for
longer  queries as  it has  been shown  that the  query term-frequency
aspect should be close to linear \cite{robertson1994}.

There are  several other disadvantages to the  {DCM-L-T} method. While
the complexity of  most retrieval functions is linear  with respect to
the number  of unique terms (word  types) in common to  both query and
document,  the complexity of  the approach  by \citet{xu08}  is linear
with  respect to  the  sum  of the  query-term  frequencies (i.e.  all
instances  of query-terms)  in  the document.  This adversely  affects
retrieval time. Conversely, the SPUD  model outlined in this work is 
as efficient at  query time  as the multinomial language model.

In the {DCM-L-T}  approach, the estimation of the  parameters for both
the  relevant and  the non-relevant  DCM document  models do  not have
closed-form  expressions.  This  is  not  of  major  concern  for  the
estimation of a non-relevant model in a static collection\footnote{For
  a dynamically changing collection where new documents are discovered
  and indexed frequently, this may  become an issue.} (which can often
be estimated off-line), but is  a major disadvantage for the inference
of  the relevant  model,  which  must be  estimated  on-line at  query
time. In fact, one of the major difficulties with the previous 
relevance-based approach is estimating the set of pseudo relevant documents 
needed in order to infer the relevance model. Therefore,  a number  of  
computationally expensive  estimation techniques are compared in order
to find parameters that are the most effective in terms of retrieval. 
However, it was found that a manual tuning of $\gamma$ is more effective 
than any of these estimation techniques.

\subsection{Estimating Free Parameters}

In Section~6.4 we have shown that both SPUD retrieval methods are more
robust  in   terms  of  parameter  settings   than  their  multinomial
counterparts.  We have  shown  that for  the  SPUD$_{dir}$ model,  the
background model  is weighted approximately  four times more  than the
document  model, and that  this setting  (via $\omega=0.8$)  is robust
across different collections. More extensive research would need to be
conducted  to  determine if  this  setting  is  universal. Some  prior
research  into  Microblog  retrieval  suggests that  a  smaller  $\mu$
parameter  value in  the multinomial  query likelihood  model  is more
effective    on   collections    that   contain    smaller   documents
\cite{han12,kim12}. This is consistent with our results (emphasised by
results on the \emph{ohsu}  collection which contains short documents)
as  the estimate  of $m_c$  is  correlated with  document length  (see
Table~\ref{tab:rbow_dir_optimal}). This provides further evidence that
our  free  hyperparameter  $\omega$  is  more  robust  than  the  free
parameter $\mu$ in the multinomial model.

Furthermore,  although it  has been  suggested \cite{zhai04}  that the
parameter  $\mu$ in  the original  multinomial language  model  may be
affected by query  length, we have found that  the most effective SPUD
retrieval method  is robust across  queries of different  length. More
work  would need  to  be conducted  to  see if  the  optimal value  of
$\omega$   varies   according  to   query   length.  Recent   research
\cite{tsagkias11}  has investigated a  different generative  model for
queries, and  this would  also be an  interesting future  direction to
explore.

The background model in SPUD is only an efficient approximation to the
DCM.  Although,  it  has  been  shown \cite{elkan06}  that  this  EDCM
approximation is  quite accurate and has  been shown to  be useful for
text clustering,  more extensive  work would need  to be  conducted to
determine  if  the approximation  is  close  to  optimal in  terms  of
retrieval effectiveness.

\subsection{Theoretical Discussions}

\subsubsection{Term and Document Event Spaces}

Aspects  of both  term-frequency and  inverse document  frequency have
been  at  the core  of  many  successful  ranking functions  over  the
years.  The work  outlined here  helps to  explain why  both  of these
features  have   been  so   useful.  In  particular,   the  generative
assumptions made in our document model help explain why term-frequency
is such a useful and salient measure of topicality. In other words, we
argue that it is because  authors have preferential attachment for the
content  words  within-documents that  term-frequency  is such  useful
measure of topicality.  Furthermore, these generative assumptions lead
to    power-law   characteristics    of    term-frequency   in    text
\cite{simon55,goldwater11}, and therefore  appear to be more plausible
models.

Interestingly,  it is  because  of within-document  \emph{preferential
  attachment}  that inverse  document  frequency is  such an  accurate
measure   of  term-specificity.   Essentially,   when  analysing   the
collection-wide characteristics  of terms, for  the most part  we need
only count  the first occurrence of  a term within a  document, as all
other occurrences depend upon this. While  we did not derive idf as it
appeared in its original  form \cite{jones72}, our analysis shows that
the  best retrieval  formula  derived from  the  SPUD language  model,
contains  characteristics   closely  related  to  that   of  idf  (see
Fig.~\ref{fig:idf}). By capturing burstiness  in our framework we have
been able  to successfully  combine the term  event space  used within
each   document,  with   the  document   event  space   used   at  the
collection-wide level  (which comes about as a  close approximation to
the background DCM).  Others \cite{robertson04,roelleke08} have argued
that Harter's  \emph{eliteness} hypothesis \cite{harter75a,harter75b},
which is  essentially a binary latent  variable for each  term, acts a
\emph{bridge} between the  term space and the document  space. We have
found that  there are  alternative generative explanations  for tf-idf
type schemes. We believe that  the SPUD language model is an important
step towards  developing a probabilistic  generative theory explaining
such schemes.

\subsubsection{Relevance} 
We note  that our  retrieval model is  a query-likelihood  model which
does not  explicitly model relevance;  however it is not  difficult to
place  the  same document  model  in  a  relevance framework.  The  KL
(Kullback-Leibler)   divergence,   which   measures  the   amount   of
information  lost  when one  distribution  is  used  to model  another
theoretical distribution,  has been  used in information  retrieval to
compare document models  to query models. As this  introduces the idea
of a query model, it seems reasonable to imagine that this query model
is  a best initial  approximation of  the \emph{true}  relevance model
(which   can    be   updated   as    relevance   information   becomes
known). Therefore, one can  think about ranking documents according to
the negative  KL-divergence of a document model  $\mathcal{M}_d$ and a
true relevance model $\mathcal{M}_r$ as follows:

\begin{equation} \label{eq:kl}
- KL(\mathcal{M}_r || \mathcal{M}_d) =  - \sum_{t \in v} p(t|\mathcal{M}_r) \cdot log\frac{p(t|\mathcal{M}_r)}{p(t|\mathcal{M}_d)}
\end{equation}
It  is also  well-known  that the  query-likelihood  function is  rank
equivalent  to the KL-divergence  between a  query model  and document
model as a special case \cite{zhai01:mf,zhai08}. The above equation is
rank    equivalent   to    the   SPUD    retrieval    functions   when
$p(t|\mathcal{M}_r)$   is  estimated   using  $c(t,q)/|q|$   and  when
$p(t|\mathcal{M}_d)$  is  estimated  using  the  new  document  models
presented in this article (i.e. $\bm{\alpha}_{dm}$).

\subsubsection{Document Length Normalisation}
In  Section~5.2.1 we  defined a  constraint to  capture  the verbosity
hypothesis.  We have  shown that  the best  performing  SPUD retrieval
method adheres  to this constraint. We  have seen that  in general the
multinomial  model MQL$_{dir}$ over-penalises  long documents  and the
SPUD$_{dir}$ model  is more likely  to retrieve longer  documents (See
Fig.~\ref{fig:prob_rel}). This  is because the  multinomial model does
not  model the  distinction  between word-types  and word-tokens,  and
ultimately  over-penalises  documents  with recurrences  of  non-query
terms.  This result builds  on recent  research \cite{lv11,cummins12d}
that   developed  further   constraints   regarding  document   length
normalisation. It would be interesting future research to determine if
the SPUD$_{dir}$ function adheres to these constraints also.
 
The SPUD  model significantly  outperforms a highly  tuned multinomial
model  (MQL) for all  query lengths.  This is  because the  SPUD model
incorporates two types of document length normalisation. One aspect of
normalisation (verbosity) regulates the term-frequency with respect to
the document length as longer  documents (those with many word tokens)
are more likely to contain higher term-frequencies. The another aspect
(scope) normalises  longer documents (those  with more word  types) as
they are more likely to contain more distinct query-terms. This second
aspect of  normalisation is crucially  dependant on query  length. The
SPUD model is the first model to combine these two aspects of document
normalisation in a theoretically principled framework.

Interestingly,  recent  research has  developed  a two-stage  document
length normalisation  framework \cite{na2015} which  incorporates both
verbosity  and  scope  normalisation  into retrieval  methods.  It  is
appealing   that  the   SPUD  retrieval   methods  derived   from  our
probabilistic framework contain these aspects of normalisation naturally.

\subsection{Broader Impact}

While we  have argued that  the new SPUD  model addresses a  number of
theoretically interesting  questions in IR, we  have demonstrated that
it also  practically useful  in a retrieval  scenario. Given  that the
SPUD  model  is  essentially   a  method  for  determining  principled
term-weights for document vectors, the model is likely to be useful in
other areas  where term-weights are used in  vector representations of
longer texts.  This includes areas  such as text  classification, text
clustering, and  more specialised NLP tasks  (e.g. keyword extraction,
automatic summarisation).

\subsection{Recommended Retrieval Function}

The    recommended    retrieval    function   is    SPUD$_{dir}$    in
Eq.~(\ref{eq:rbow_dir2}). This function  has one free parameter $\mu'$
which we recommend setting to $4  \cdot m_c$, where $m_c$ can be found
by applying Newton's  method to Eq.~(\ref{eq:estimate_w}). Alternative
$\mu'$  can be  experimentally tuned  on  training data  which is  the
current method of setting $u$ in the multinomial language model.

\subsection{Future Directions}

The     most     effective      SPUD     method,     introduced     in
Eq.~(\ref{eq:rbow_mix1}), linearly  combines the background  DCM model
with  the document  DCM  language  model. This  could  be extended  to
include  more language  models.  For example,  if  we had  information
relating  to authorship,  we  could estimate  an  author specific  DCM
language model that would  explain textual characteristics specific to
an author,  as it may be  the case that certain  authors are generally
more  verbose than  others. Smoothing  this  DCM model  with both  the
document and  background models may further  improve performance. This
may be particular useful in areas such as expert search.

The document model  outlined in this work models  word burstiness in a
document specific  manner. Previous work \cite{kwok96}  has shown that
certain terms are  more bursty than others (i.e.  they are more likely
to repeat). This suggests that incorporating a term-specific aspect of
burstiness  may increase  retrieval effectiveness  even  further. This
could be  modelled using a more  general urn model where  the level of
reinforcement varies per term.

A  further  interesting  direction  is  to  consider  integrating  the
document  model  outlined here  with  a  model  that incorporates  the
traditional notion  of term-dependence.  The details regarding  such a
combination have not been discussed here but would present interesting
future research.

%On \emph{trec-8} data, $\delta$ approximately ranges between $0.05$ and $0.5$
% for typical query-terms. \citet{sunehag07} has previously outlined a similar argument.

\section{Conclusion}

We have introduced a new  family of language model (namely SPUD) based
on  a P\'olya  urn  process. We  have  shown that  a query  likelihood
retrieval  method based  on  this model  is  superior to  that of  the
state-of-the-art  multinomial language  model. Interestingly,  we have
shown  that  the new  model  can be  computed  as  efficiently as  the
multinomial  language model.  Essentially,  this means  that the  SPUD
retrieval  method  can be  used  in  place  of the  multinomial  query
likelihood  method  in   many  different  retrieval  applications  and
domains.

We have outlined a number of  intuitions that help to motivate the new
model.   For example,  we  developed a  constraint  for the  verbosity
hypothesis and  have shown  that the most  effective SPUD  method, the
SPUD$_{dir}$ model, adheres to  this constraint.  Furthermore, we have
shown  that  the  free  hyperparameter  (i.e.   $\omega=0.8$)  in  the
SPUD$_{dir}$  method  is  robust  across  various  collections.   This
essentially  reduces  the need  for  experimental  tuning.  Given  the
principled  nature of  the approach  developed, it  can be  used  in a
variety of  IR tasks. We have  shown that it is  useful for downstream
retrieval methods, as  we have used it to  estimate a pseudo-relevance
based model (PURM)  that demonstrates improved retrieval effectiveness
on test collections when compared to a pseudo-relevance model based on
the multinomial (RM3).

%In this article we have modelled only document specific burstiness. We 
%can extend the model further to capture author specific characteristics 
%based on documents written by an author. This extended model may potentially 
%boost retrieval effectiveness and may be particularly useful in scholarly domains 
%and for the expert search task.

Future  work   will  look   to  improve  retrieval   effectiveness  by
incorporating multiple  DCM language models for  modelling a document.
Furthermore, we  aim to investigate the query  likelihood method using
different  generative assumptions  for  the query.  In  this work,  we
assumed  a sampling-with-replacement  strategy  for query  generation.
However,  different sampling  strategies,  such as  those employed  by
Friedman urn's \cite{friedman65} might better model query generation.

\section*{Acknowledgements}
We would  like to  thank the anonymous  reviewers whose  comments have
greatly helped to improve and clarify this work.

\bibliographystyle{abbrvnat}
\bibliography{references}

\end{document}